\documentclass[pra,floatfix,twocolumn,showpacs]{revtex4}
\usepackage{amsmath,amsfonts,amssymb}
\usepackage{graphicx}

\renewcommand{\vec}{\boldsymbol}
\renewcommand{\Re}{\text{Re}\,}
\renewcommand{\Im}{\text{Im}\,}

\begin{document}

\title{Closed orbits and their bifurcations in the crossed-fields hydrogen
       atom}
\author{Thomas Bartsch}
\author{J\"org Main}
\author{G\"unter Wunner}
\affiliation{Institut f\"ur Theoretische Physik 1, Universit\"at Stuttgart,
         D-70550 Stuttgart, Germany}
\date{\today}

\begin{abstract}
A systematic study of closed classical orbits of the hydrogen atom in
crossed electric and magnetic fields is presented. We develop a local
bifurcation theory for closed orbits which is analogous to the well-known
bifurcation theory for periodic orbits and allows identifying the generic
closed-orbit bifurcations of codimension one. Several bifurcation scenarios
are described in detail. They are shown to have as their constituents the
generic codimension-one bifurcations, which combine into a rich variety of
complicated scenarios. We propose heuristic criteria for a classification
of closed orbits that can serve to systematize the complex set of orbits.
\end{abstract}
\maketitle

\section{Introduction}
\label{chap:Intro}

Closed-orbit theory \cite{Du88,Bogomolny89} has proven to be the key tool
to analyze the photoabsorption spectra of atoms in external fields. It
interprets spectral oscillations semiclassically in terms of closed orbits
of the underlying classical system, i.e. of classical orbits starting at
and returning to the nucleus. A complete semiclassical description of an
atomic spectrum therefore requires a sufficiently detailed understanding of
the classical closed orbits. In particular, the possible types of
closed-orbit bifurcations must be described, so that the generation of new
closed orbits upon varying the external field strengths can be accounted
for.

For the hydrogen atom in a magnetic field, the systematics of closed orbits
and their bifurcations has been known for a long time
\cite{Main91,Main86,AlLaithy86,AlLaithy87,Main87,Mao92}. For the hydrogen
atom in crossed electric and magnetic fields, the classical mechanics is
much more complicated because three non-separable degrees of freedom have
to be dealt with. Although a number of closed orbits have been identified
in experimental or theoretical quantum spectra
\cite{Main91,Raithel91,Raithel94,Rao01,Freund02}, a systematic study of
these orbits and their bifurcations is still lacking.

Considerable effort has been spent during the past decade on the study of
the classical mechanics of the crossed-fields hydrogen atom in the limit of
weak external fields
\cite{Gourlay93,Floethmann94,Milczewski97b,Sadovskii98,Berglund00}. That
work relies on the observation that for weak external fields the principal
quantum number $n$, or its classical analogue $n=1/\sqrt{-2E}$, is conserved
to a higher degree of precision than the angular-momentum and Lenz vectors
$\vec L$ and $\vec A$. The latter are conserved in the pure Kepler problem,
but acquire a slow time-dependence in weak fields, so that the electron can
be visualized as moving on a slowly precessing Kepler ellipse.

The most important result in the present context is the finding first
described in \cite{Floethmann94} that there are four Kepler ellipses that
remain unperturbed by the external fields to first order in the field
strength, i.e. among the continuous infinity of periodic orbits of the
unperturbed Kepler proble there are four orbits that are periodic even in
the presence of external fields. These fundamental periodic orbits can be
regarded as the roots of ``family trees'' of periodic orbits. More
complicated orbits are created out of the fundamental orbits by
bifurcations as the field strengths increase.

However, none of the fundamental periodic orbits is closed at the
nucleus. Their knowledge therefore does not aid in the classification of
closed orbits. A first systematic study of closed orbits in the
crossed-fields system and their bifurcations was performed by Wang and
Delos \cite{Wang01}. These authors presented orderly sequences of
bifurcations of planar closed orbits (i.e.  orbits in the plane
perpendicular to the magnetic field), which they interpreted in terms of an
integrable model Hamiltonian.

In the present paper we undertake a systematic investigation of closed
orbits and their bifurcations in the crossed-fields hydrogen atom.  In
section~\ref{sec:classHam}, the symmetries of the Hamiltonian are briefly
reviewed. Section~\ref{sec:BifGen} presents the general framework of a
local bifurcation theory of closed orbits, and section~\ref{sec:Cod1Bif}
describes the generic codimension-one bifurcations. A discussion of
complex ghost orbits is included in each case because they are known to play
an important role in semiclassics \cite{Kus93,Main97a}. In
section~\ref{sec:ClosedDKP}, the closed orbits in the hydrogen atom in a
magnetic field are surveyed.  Section~\ref{sec:XCO} then
details the bifurcation scenarios actually observed in the crossed-fields
system.  It is shown that the elementary codimension-one bifurcations
actually form the building blocks of the bifurcations scenarios, but that
in many cases complicated scenarios consisting of several elementary
bifurcations occur.  In section~\ref{sec:class}, a heuristic classification
scheme for the closed orbits in crossed fields is proposed which is
based on the well-known classification for the closed orbits in a magnetic
field. The actual semiclassical quantization of the crossed-fields hydrogen
atom in the framework of closed-orbit theory, which is based on the results
presented here, is described in an accompanying paper \cite{Bartsch03b}.

\section{The classical Hamiltonian}
\label{sec:classHam}

Throughout this work, we will assume the magnetic field to be directed
along the $z$-axis and the electric field to be directed along the
$x$-axis. In atomic units, the Hamiltonian describing the motion of the
atomic electron then reads
\begin{equation}
  \label{specHam}
  H=\frac12\vec p^2-\frac 1r + \frac 12BL_z+\frac 18B^2\rho^2+Fx \;,
\end{equation}
where $r^2=x^2+y^2+z^2$, $\rho^2=x^2+y^2$ and $L_z$ is the $z$-component of
the angular momentum vector.  By virtue of the scaling properties of the
Hamiltonian~(\ref{specHam}), the dynamics does not depend on the energy $E$
and the field strengths $B$ and $F$ separately, but only on the scaled
energy $\tilde E=B^{-2/3}E$ and the scaled electric field strength $\tilde
F=B^{-4/3}F$. Upon scaling, all classical quantities are multiplied by
suitable powers of the magnetic field strength $B$. In particular,
classical actions scale according to $\tilde S=B^{-1/3}S$. These scaling
prescriptions will be used throughout this work.

In crossed fields, two angles are required to characterize the
starting or returning direction of a closed orbit. We will use the polar
angle $\vartheta$ between the trajectory and the magnetic field axis and
the azimuthal angle $\varphi$ between the projection of the trajectory into
the $x$-$y$-plane and the electric field axis.

The hydrogen atom in crossed fields does not possess any continuous
symmetries so that, apart from the energy, no constant of the motion exists
and three non-separable degrees of freedom have to be dealt with.  There
are, however, three discrete symmetry transformations of the crossed-fields
system, namely
\begin{itemize}
\item the reflection \textsf{Z}\ at the $x$-$y$-plane,
\item the combination \textsf{T}\ of time-reversal and a reflection at the
$x$-$z$-plane,
\item the combination \textsf{C}=\textsf{Z}\textsf{T}\ of the above.
\end{itemize}
The effects of the transformations on
the initial and final angles of a closed orbit are summarized in
table~\ref{tab:SymmAng}.

\begin{table}
  \begin{center}
  \begin{tabular}{c|crcr|c}
    & \multicolumn{4}{c|} {Transformation} & Symmetry conditions\\ 
        & $\vartheta_i$ & $\varphi_i$ & $\vartheta_f$ & $\varphi_f$ 
        & \\ \hline
    \textsf{Z}  & $\pi-\vartheta_i$ & $\varphi_i$ & $\pi-\vartheta_f$ & $\varphi_f$
        & $\vartheta_i=\vartheta_f=\frac{\pi}{2}$ \\
    \textsf{T}  & $\vartheta_f$ & $-\varphi_f$ & $\vartheta_i$ & $-\varphi_i$
        & $\vartheta_i=\vartheta_f$ and $\varphi_i=-\varphi_f$ \\
    \textsf{C}  & $\pi-\vartheta_f$ & $-\varphi_f$ & $\pi-\vartheta_i$ & $-\varphi_i$
        & $\vartheta_i=\pi-\vartheta_f$ and $\varphi_i=-\varphi_f$
  \end{tabular}
  \end{center}
  \caption{\label{tab:SymmAng}
    The symmetry transformations of the crossed-fields system:
    Transformation of initial and final angles and symmetry conditions for
    doublets. Singlets satisfy $\vartheta_i=\vartheta_f=\frac{\pi}{2}$
                       and $\varphi_i=-\varphi_f$.}
\end{table}

The application of these transformations to a given closed orbit yields a
group of four closed orbits of equal length. Typically, these orbits will
all be distinct, so that closed orbits in the crossed-fields system occur
in quartets. In particular cases, a closed orbit can be invariant under one
of the symmetry transformations. In this case, there are only two distinct
orbits related by symmetry transformations. We will refer to them as a
doublet, or more specifically as a \textsf{Z}-, \textsf{T}-, or
\textsf{C}-doublet, giving the transformation under which the orbits are
invariant. The transformations of the initial and final angles listed in
table~\ref{tab:SymmAng} yield symmetry conditions that an orbit invariant
under any of the transformations must satisfy. These are also given in
table~\ref{tab:SymmAng}. In special cases, a closed orbit can be invariant
under all three symmetry transformations. It
then occurs as a singlet, since no distinct orbits can be generated from it
by a symmetry transformation.

Among the symmetry transformations, the reflection \textsf{Z}\ plays a
special role in that it is a purely geometric transformation. There is,
therefore, an invariant subspace of the phase space, viz. the $x$-$y$-plane
perpendicular to the magnetic field. This plane is invariant under the
dynamics and therefore forms a subsystem with two degrees of freedom.

In connection with bifurcations of orbits it is essential for semiclassical
applications to study complex ``ghost'' orbits along with the real orbits,
i.e. to allow coordinates and momenta to assume complex values. For ghost
orbits, another reflection symmetry arises, viz. the symmetry with respect
to complex conjugation. Since the Hamiltonian~(\ref{specHam}) is real, it
is invariant under complex conjugation. Therefore, ghost orbits always
occur in pairs of conjugate orbits.

\section{Closed-orbit bifurcation theory}

\subsection{General theory}
\label{sec:BifGen}

The dynamics of the hydrogen atom in a pure magnetic field
possesses time-reversal invariance if it is restricted to the subspace of
vanishing angular momentum $L_z$. Due to this symmetry, an electron
returning to the nucleus will rebound from the Coulomb center into its
direction of incidence and retrace its previous trajectory back to its
starting direction. Therefore, any closed orbit is either itself periodic
or it is one half of a periodic orbit. Due to the close link between closed
orbits and periodic orbits, closed-orbit bifurcations can be described in
the framework of periodic-orbit bifurcation theory developed by Mayer
\cite{Mayer70,Mao92}.  In particular, in a magnetic field closed orbits
possess repetitions, so that arbitrary $m$-tupling bifurcations are
possible.

In the presence of crossed electric and magnetic fields, the time-reversal
invariance is broken, and no general connection between closed orbits and
periodic orbits remains. As a consequence, the techniques of periodic-orbit
bifurcation theory are no longer applicable, and a novel approach to the
classification of closed-orbit bifurcations must be found. In this section,
a general framework for the discussion of closed-orbit bifurcations will be
introduced.

The crucial step in the development of the bifurcation theory of periodic
orbits is the introduction of a Poincar\'e surface of section map in the
neighborhood of the orbit. The Poincar\'e map describes the dynamics of
the degrees of freedom transverse to the orbit, and the orbit bifurcates
when the transverse dynamics becomes resonant with the motion along the
orbit.

For periodic orbits, a Poincar\'e map is specified by fixing a surface of
section in phase space which is transverse to the orbit. For a point $P$ on
the surface of section, the trajectory starting at $P$ is followed until it
intersects the surface of section again. This intersection point is defined
to be the image of $P$ under the Poincar\'e map. The periodic orbit itself
returns to its starting point, so that it appears as a fixed point of the
Poincar\'e map.

This prescription is not directly applicable to closed orbits because they
do not return to their starting point in phase space. Therefore, a
trajectory starting on the surface of section will not in general intersect
the surface again. To arrive at a meaningful definition of a Poincar\'e
map, one must use two surfaces of section: the first transverse to the
initial direction of the orbit, the second transverse to its final
direction. A trajectory starting in the neighborhood of the closed orbit on
the initial surface of section $\Sigma_i$ will then have an intersection
with the final section $\Sigma_f$, so that a Poincar\'e map is well
defined. As in the case of a periodic orbit, the Poincar\'e map is
symplectic.

Unlike with periodic orbits, the notion of a closed orbit is not invariant
under canonical transformations. The distinction between position space and
momentum space must therefore be kept. Let $(q_i,p_i)$ and $(q_f,p_f)$ be
canonical coordinates on the surfaces $\Sigma_i$ and $\Sigma_f$ chosen so
that $q_i$ and $q_f$ are position space coordinates in the directions
perpendicular to the initial or final directions of the orbit. The origins
of the coordinate systems are fixed so that the position of the nucleus is
$q_i=0$ or $q_f=0$, respectively. Closed orbits are then characterized by
$q_i=q_f=0$. In crossed fields three spatial dimensions must be dealt with,
so that each of $q_i,p_i,q_f,p_f$ is a two-dimensional vector.  The reader
may conveniently picture $q_i$ and $q_f$ as Cartesian coordinates, although
in this case the conjugate momenta $p_i$ and $p_f$ diverge as the Coulomb
singularity is approached. This difficulty can be overcome by means of a
Kustaanheimo-Stiefel regularization \cite{KS65}. Coordinates having the
properties described above can then be shown to exist, as will be discussed
in detail elsewhere \cite{Bartsch02,Bartsch03c}.

A closed orbit can start in $\Sigma_i$ with arbitrary initial momentum
$p_i$, but it must start in the plane $q_i=0$. The Poincar\'e map maps this
plane into a Lagrangian manifold in $\Sigma_f$. Closed orbits are given by
the intersections of this manifold with the plane $q_f=0$. In a less
geometrical way of speaking, closed orbits can be described as solutions of
the equation $q_f(p_i,q_i=0) = 0$. A particular solution of this equation,
corresponding to the orbit the construction started with, is given by
$q_f(p_i=0)=0$. If the matrix $B=\partial q_f/\partial p_i$ is non-singular
at $p_i=0$, this solution is locally unique and, by the implicit function
theorem, will persist upon the variation of parameters. Thus, the closed
orbit cannot undergo a bifurcation unless $M=\det B=0$.

\begin{table}
  \begin{tabular}{c|cc|c}
    type & \multicolumn{2}{c|}{transformation} & regular matrix \\
    \hline
    $F_1(q_i,q_f)$ & $p_i=+\partial F_1/\partial q_i$\;, &
                     $p_f=-\partial F_1/\partial q_f$ & $B$ \\
    $F_2(q_i,p_f)$ & $p_i=+\partial F_2/\partial q_i$\,, &
                     $q_f=+\partial F_2/\partial p_f$ & $D$ \\
    $F_3(p_i,q_f)$ & $q_i=-\partial F_3/\partial p_i$\,, &
                     $p_f=-\partial F_3/\partial q_f$ & $A$ \\
    $F_4(p_i,p_f)$ & $q_i=-\partial F_4/\partial p_i$\,, &
                     $q_f=+\partial F_4/\partial p_f$ & $C$
  \end{tabular}
  \caption{Overview of generating functions of different types
  (cf.~\cite{Goldstein}).}
  \label{GenFctTab}
\end{table}

An overview of the bifurcation scenarios to be expected when $\det B=0$ can
be obtained from a description of the possible modes of behavior of the
Poincar\'e map. This can most conveniently be achieved if the Poincar\'e
map is represented by a generating function \cite{Goldstein}. The
generating function can be chosen to depend on any combination of initial
and final positions and momenta, as long as they form a complete set of
independent coordinates. We adopt the well-known conventions of
Goldstein \cite{Goldstein} for denoting different types of generating
functions, which are summarized in table~\ref{GenFctTab}.

For a generic symplectic map, all possible sets of coordinates and momenta
are independent, so that generating functions of any type exist. At a
closed-orbit bifurcation, however, a degeneracy indicated by the condition
that $B=\partial q_f/\partial p_i$ be singular arises, so that care must be
taken in choosing a generating function. Loosely speaking, if $B$ is
singular, $p_i$ cannot be determined from $q_i$ and $q_f$, so that it may
be conjectured that no generating function of type $F_1$ exists. To confirm
this conjecture, we study a linear symplectic map
\begin{equation}
  \label{LinSymp}
  q_f = A q_i + B p_i \;, \qquad p_f = C q_i + D p_i
\end{equation}
with four matrices $A, B, C, D$ satisfying the symplecticity conditions
\cite{McDuff95}
\begin{equation}
  \label{SympCond}
  \begin{alignedat}{3}
    A^\top C &= C^\top A \;, \quad B^\top D &= D^\top B \;, \quad
    A^\top D - C^\top B &= 1\,, \\
    A B^\top &= B A^\top \;, \quad C D^\top &= D C^\top \;, \quad
    A D^\top - B C^\top &= 1 \;,
  \end{alignedat}
\end{equation}
where $^\top$ denotes the transpose.
A generating function for the linear map (\ref{LinSymp}) must be quadratic
in its variables. From the ansatz
\begin{equation}
  \label{F1Ansatz}
  F_1(q_i,q_f) = \frac 12 q_f^\top R q_f + q_f^\top S q_i 
               + \frac 12 q_i^\top T q_i
\end{equation}
with matrices $R, S, T$, the map (\ref{LinSymp}) is obtained if
\begin{equation}
  \label{F1Sol}
  \begin{gathered}
    R=-DB^{-1} \;, \qquad T=-B^{-1\top}A \;, \\
    S=B^{-1\top}=DB^{-1}A-C \;.
  \end{gathered}
\end{equation}
The two expressions given for $S$ are equal by virtue
of~(\ref{SympCond}). As expected, a generating function of type $F_1$ does
not exist if $B$ is singular.  A similar calculation can be carried out for
the other types of generating functions. For each type, one of the matrices
$A,B,C,D$ must be non-singular. These results are given in
table~\ref{GenFctTab}. Locally, they can be extended to non-linear maps by
means of the implicit function theorem.

Thus, at a bifurcation of closed orbits the Poincar\'e map generically
possesses generating functions of all types except $F_1$. The most
convenient choice is a function of type $F_4(p_i,p_f)$. The transformation
equations associated with this type of generating function read
\begin{equation}
  \label{F4Transform}
  q_i=-\frac{\partial F_4}{\partial p_i} \;, \qquad
  q_f=+\frac{\partial F_4}{\partial p_f} \;.
\end{equation}
Closed orbits are characterized by $q_i=q_f=0$. They therefore agree with
the stationary points of the $F_4$ function. The classification problem of
closed-orbit bifurcation theory can thus be rephrased as the problem to
determine how stationary points of a real function change upon the
variation of parameters. This question is the subject of catastrophe theory
\cite{Poston78,Saunders80,Castrigiano93}.

Catastrophe theory studies smooth real-valued functions $f(\vec x)$ and
$\tilde f(\vec x)$ defined in a neighborhood of the origin in an
$n$-dimensional configuration space. They are said to be equivalent if
there is a diffeomorphism $\psi(\vec x)$ of the configuration space so that
\begin{equation}
  \label{CatastEq}
  \tilde f(\vec x) = f(\psi(\vec x)) \;.
\end{equation}
The coordinate transformation $\psi$ maps the stationary points of $\tilde
f$ to those of $f$. In this sense, the distributions of stationary points
of $f$ and $\tilde f$ agree qualitatively. Without loss of generality it
can be assumed that $f$ and $\tilde f$ have stationary points at the
origin, because any stationary point can be moved there by a coordinate
transformation. After adding a constant, one has $f(0)=0$.

$f$ is said to be structurally stable if any small perturbation $\tilde f$
of $f$ (i.e. $\tilde f(\vec x) = f(\vec x)+\epsilon g(\vec x)$ with a
smooth function $g(\vec x)$ and sufficiently small $\epsilon$) is
equivalent to $f$. Notice that catastrophe theory is a purely local
theory. It is concerned with the structural stability or instability of a
single stationary point and the pattern of stationary points that can be
generated from a structurally unstable stationary point by a small
perturbation.

\begin{figure}
  \includegraphics[width=.47\columnwidth]{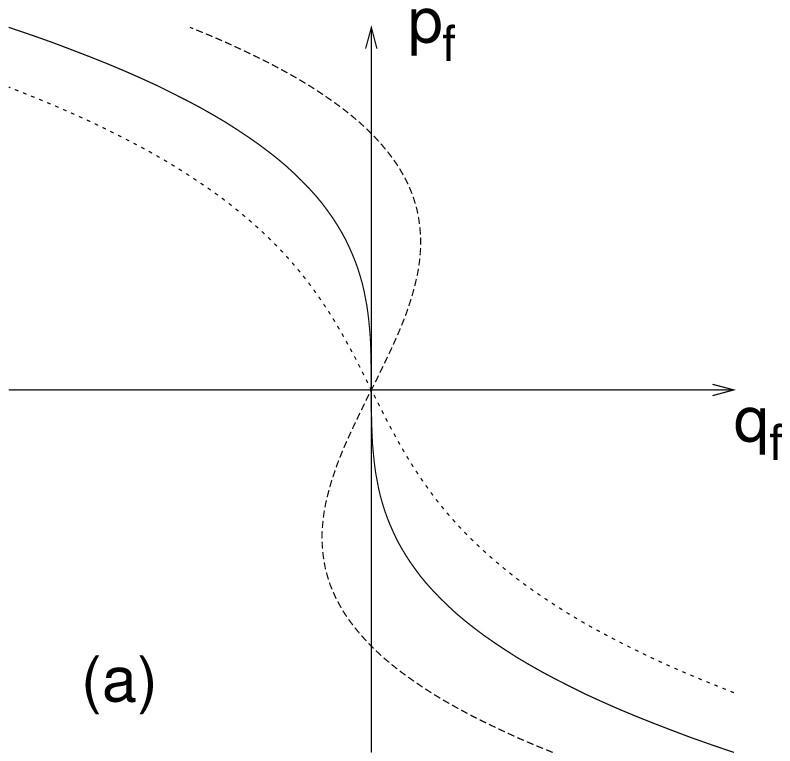}\hfill
  \includegraphics[width=.47\columnwidth]{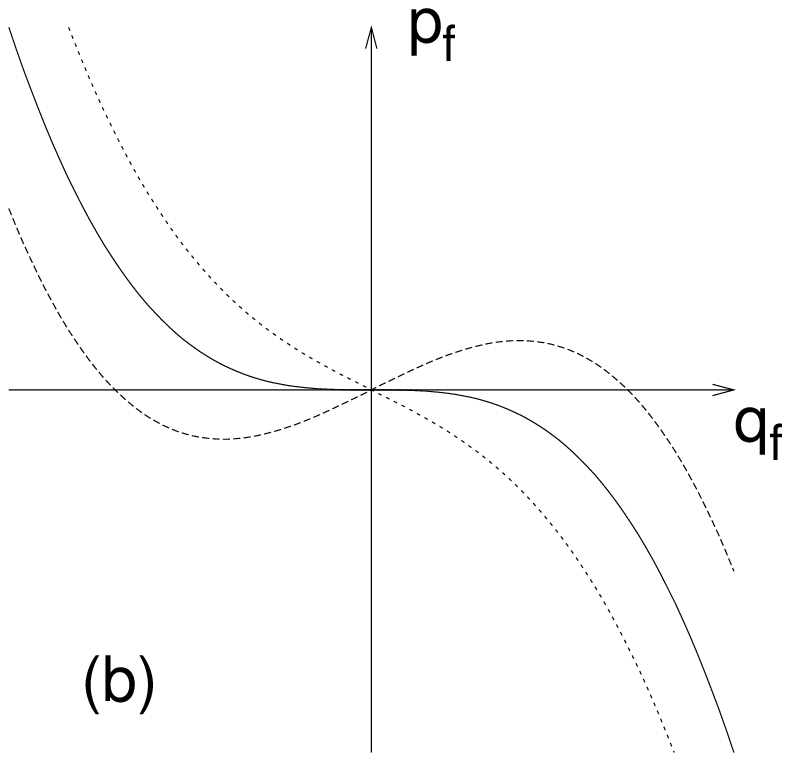}
  \caption{Schematic plot of the Lagrangian manifold $q_i=0$ in $\Sigma_f$
  for the case (a) $B=\partial q_f/\partial p_i=0$ and (b) $D=\partial
  p_f/\partial p_i=0$. The dashed lines indicate the position of the
  manifold at slightly varied parameter values. Only in case (a)
  additional intersections with the plane $q_f=0$ can arise.}
  \label{DegenFig}
\end{figure}

In the present context, non-bifurcating closed orbits correspond to
structurally stable stationary points of $F_4$, because a small variation
of parameters will bring about a variation of $F_4$ which is small in the
above sense and preserves the stationary point.  The most elementary result
of catastrophe theory states that a stationary point of a function $f$ is
structurally stable if its Hessian matrix, i.e. the matrix of second
derivatives of $f$, is non-singular. For the linear symplectic
transformation (\ref{LinSymp}), the $F_4$ generating function is
\begin{equation}
  F_4 = \frac 12 p_f^\top AC^{-1} p_f - p_f^\top C^{-1\top} p_i
       +\frac 12 p_i^\top C^{-1}D p_i \;,
\end{equation}
so that its Hessian determinant at $p_i=p_f=0$ can be found to be
\begin{equation}
  \det\operatorname{Hess} F_4 = \frac{\det B \det D}{\det C} \;.
\end{equation}
The Hessian matrix of $F_4$ is thus singular if either $B=\partial
q_f/\partial p_i$ or $D=\partial p_f/\partial p_i$ is.  It has been
shown above that bifurcations of closed orbits can only occur if $\det B=0$,
i.e. a bifurcating orbit corresponds to a degenerate stationary point of
$F_4$. The case $\det D=0$ also leads to a degeneracy of $F_4$, but it
cannot be associated with a closed-orbit bifurcation. This can also be
understood geometrically: As illustrated in figure~\ref{DegenFig}, if $\det
B=0$, the Lagrangian manifold given by $q_i=0$ is tangent to the plane
$q_f=0$, so that it can develop further intersections with that plane upon
a small variation of parameters. If $\det D=0$, the manifold is tangent to
the plane $p_f=0$, whence, upon a variation of parameters, it can acquire
additional intersections with that plane, but not with the plane $q_f=0$,
so that no bifurcation of closed orbits can arise.

The discussion of stationary points with degenerate Hessian matrices, also
called ``catastrophes'', is simplified considerably by the splitting lemma
of catastrophe theory~\cite{Castrigiano93}. It states that if the dimension
of the configuration space is $n$ and a function $f$ on the configuration
space has a stationary point at the origin whose Hessian matrix has rank
$n-m$, a coordinate system $x_1,\dots,x_n$ can be introduced in a
neighborhood of the stationary point so that
\begin{equation}
  \label{splitting}
  f(x_1,\dots,x_n) = g(x_1,\dots,x_m) + q(x_{m+1},\dots,x_n) \;,
\end{equation}
where $q$ is a non-degenerate quadratic form of $n-m$ variables and the
function $g$ has a stationary point with zero Hessian matrix at the
origin. As the non-degenerate stationary point of $q$ is structurally
stable, the behavior of the stationary points of $f$ under a small
perturbation is determined by $g$ only. The number of relevant variables is
thus only $m$, which is called the corank of the catastrophe. It will be
assumed henceforth that a splitting according to~(\ref{splitting}) has been
carried out and the non-degenerate part $q$ is ignored, so that the Hessian
matrix of $f$ vanishes at the origin.

Under a small perturbation of the function $f$, a degenerate stationary
point will in general split into several distinct stationary points. This
process will be used to model bifurcations of closed orbits. The degenerate
stationary points relevant to bifurcation theory are those of finite
codimension, i.e. those for which there are smooth functions $g_1(\vec x),
\dots, g_k(\vec x)$ so that any small perturbation of $f$ is equivalent to
\begin{equation}
  \label{unfolding}
  F(\vec x) = f(\vec x) +
     \alpha_1 g_1(\vec x) + \dots + \alpha_k g_k(\vec x)
\end{equation}
with suitably chosen constants $\alpha_i$.
The function $F(\vec x)$ is called an unfolding of $f(\vec x)$, because the
degenerate stationary point of $f$ can be regarded as a set of several
stationary points that accidentally coincide and are ``unfolded'' by the
parameters $\alpha_i$. The smallest value of $k$ that can be chosen in
(\ref{unfolding}) is called the codimension of $f$. An unfolding of $f$
with $k$ equal to the codimension of $f$ is referred to as universal.

In the bifurcation problem, the generating function $F_4$ depends on
external control parameters $\rho_1,\dots,\rho_l$ such as, e.g., the energy
$E$ or the external field strengths. If, for a critical value of the
parameters, $F_4$ has a degenerate stationary point equivalent to that of
$f$, in a neighborhood of the critical value $F_4$ is equivalent to the
unfolding (\ref{unfolding}), where the unfolding parameters $\alpha_i$ are
smooth functions of the control parameters $\rho_j$. The critical parameter
values themselves are characterized by the condition that all unfolding
parameters vanish, i.e. by the set of equations
\begin{equation}
  \label{unfEq}
  \begin{gathered}
    \alpha_1(\rho_1,\dots,\rho_l) = 0 \;, \\
      \dots \\
    \alpha_k(\rho_1,\dots,\rho_l) = 0 \;.
  \end{gathered}
\end{equation}
These are $k$ equations in $l$ unknowns. They can ``generically''
only be expected to possess a solution if $k\le l$, that is, the
codimension of the degenerate stationary point must not be larger than the
number of external parameters. This construction introduces a notion of
codimension for bifurcations of closed orbits which is entirely analogous
to the codimension of bifurcations of periodic orbits: Bifurcations of a
codimension higher than the number of external parameters cannot be
expected to occur because they are structurally unstable. Under a small
perturbation of the system they would split into a sequence of ``generic''
bifurcations of lower codimensions.

\subsection{Codimension-one generic bifurcations}
\label{sec:Cod1Bif}

The considerations of the preceding section reduce the bifurcation theory
for closed orbits to the problem of determining all catastrophes having a
codimension smaller than the number of external parameters. In particular,
it explains why only catastrophes of finite codimension are relevant. In
the crossed-fields system, the number of parameters is two, if the scaling
properties are taken into account. However, we will only describe
bifurcations of codimension one in the following. They suffice to describe
the bifurcations encountered if a single parameter is varied while the
second is kept fixed. They also give a good impression of the
codimension-two scenarios because a bifurcation of codimension two must
split into a sequence of codimension-one bifurcations as soon as any of the
parameters is changed.

For generic functions without special symmetries, a list of catastrophes of
codimensions up to six with their universal unfoldings is readily available
in the literature \cite{Poston78,Saunders80,Castrigiano93}. The
classification of closed-orbit bifurcations presented here relies on these
results.

\subsubsection{The tangent bifurcation}
\label{ssec:TangentBif}

There is a single catastrophe of codimension one, which has corank one and
is known as the fold catastrophe. Its universal unfolding is given by
\begin{equation}
  \label{Fold}
  \Phi_a(t) = \frac 13 t^3 - a t \;,
\end{equation}
with $a$ denoting the unfolding parameter. The fold has two stationary
points at
\begin{equation}
  \label{FoldStat}
  t=\pm\sqrt{a} \;,
\end{equation}
where it assumes the stationary values
\begin{equation}
  \label{FoldVal}
  \Phi_a(\pm\sqrt{a}) = \mp\frac 23 a^{3/2} \;.
\end{equation}
The second derivative in the stationary points is
\begin{equation}
  \Phi_a''(\pm\sqrt{a}) = \pm 2\sqrt{a} \;.
\end{equation}
The stationary points are real if $a>0$. If $a<0$, there are no stationary
points on the real axis, because the solutions (\ref{FoldStat}) are
imaginary. These complex stationary points correspond to closed ghost
orbits in the complexified phase space. As $a$ is varied, a tangent
bifurcation occurs at $a=0$, where two complex conjugate ghost orbits turn
into two real orbits or vice versa.

\begin{figure}
  \includegraphics[width=\columnwidth]{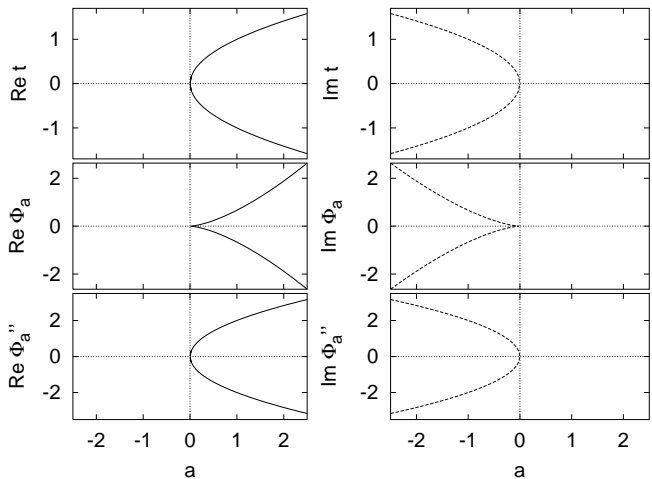}
  \caption{The positions of stationary points, stationary values and second
  derivatives in the fold catastrophe. Solid lines indicate real stationary
  points, dashed lines complex stationary points. Dotted lines are
  coordinate axes.}
  \label{FoldFig}
\end{figure}

All qualitative features of the bifurcation are described by the normal
form (\ref{Fold}). The stationary points, i.e. the closed orbits, initially
move apart as $\sqrt{a}$. A more detailed connection between the properties
of the normal form and the closed orbits can be made in the context of
uniform semiclassical approximations.  The difference between the
stationary values gives the action difference between the closed orbits,
whereas the second derivatives -- or, if the normal form has corank greater
than one, the Hessian determinants -- at the stationary points are
proportional to a parameter $M$ describing the stability of the closed
orbit (see \cite{Bartsch02,Bartsch03b} for details). All these quantities
are shown in figure~\ref{FoldFig}. When they are compared to the
corresponding quantities calculated for an actual bifurcation in
section~\ref{sec:XCO}, the qualitative agreement will become clear.

The fold catastrophe (\ref{Fold}) describes the generation of two closed
orbits in a tangent bifurcation. As this is the only generic catastrophe of
codimension one, it follows that the tangent bifurcation is the only
possible type of closed-orbit bifurcations. In particular, once it has been
generated a closed orbit cannot split into several orbits, as periodic
orbits typically do. However, this statement needs some modification due to
the presence of reflection symmetries in the crossed-fields system.

\subsubsection{The pitchfork bifurcation}
\label{ssec:PitchBif}

If the orbit under study is symmetric under
one of the reflections, i.e. it is a singlet or a doublet orbit, the
generating function $F_4$ in the neighborhood of this orbit must also
possess this reflection symmetry. By this constraint, several of the
elementary catastrophes are excluded altogether. For others, the
codimension is reduced because the unfolding can only contain symmetric
terms.

One additional catastrophe of codimension one arises, viz. the
symmetrized version of the cusp catastrophe
\begin{equation}
  \label{Cusp}
  \Phi_a(t) = \frac 14 t^4 - \frac 12 a t^2 \;.
\end{equation}
This normal form possesses the reflection symmetry $t\mapsto -t$, so that
the origin is mapped onto itself under the symmetry transformation.
There is a stationary point at the origin for all values of
the parameter $a$, corresponding to a closed orbit which is invariant under
the reflection.
Additional stationary points are located at
\begin{equation}
  \label{CuspSP}
  t=\pm\sqrt{a}\;.
\end{equation}
They are real if $a>0$ and are mapped onto each other under a
reflection. Thus, the symmetric cusp (\ref{Cusp}) describes a pitchfork
bifurcation at $a=0$, where two asymmetric orbits bifurcate off a symmetric
orbit, generating a quartet from a doublet or a doublet from a singlet.

\begin{figure}
  \includegraphics[width=\columnwidth]{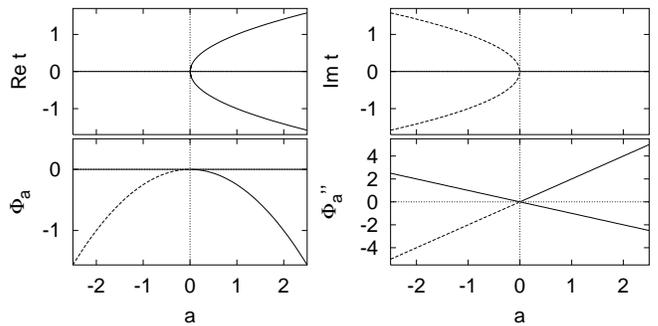}
  \caption{The positions of stationary points, stationary values and second
  derivatives in the cusp catastrophe. Solid lines indicate real stationary
  points, dashed lines complex stationary points. Dotted lines are
  coordinate axes. Note that stationary values and second derivatives are
  real even for complex stationary points.}
  \label{CuspFig}
\end{figure}

The stationary values at the asymmetric stationary points are given by
\begin{equation}
  \label{CuspVal}
  \Phi_a(\pm\sqrt{a}) = -\frac 14 a^2 \;,
\end{equation}
the second derivative is
\begin{equation}
  \Phi''_a(\pm\sqrt{a}) = 2a \;.
\end{equation}
Both the stationary values and the values of the second derivative are real
even for $a<0$, when the stationary points themselves are
complex. Therefore, these stationary points correspond to ghost orbits
having real actions and stability determinants. The existence of this
remarkable type of ghost orbits
is again a consequence of the reflection symmetry: As the stationary
points~(\ref{CuspSP}) are imaginary, the reflection $t\mapsto -t$ changes a
stationary point and its stationary value into their complex conjugates. On
the other hand, the stationary values are invariant under the reflection,
so they must be real. A ghost orbit having this symmetry property will be
referred to as a symmetric ghost orbit.

The characteristic quantities of the symmetric cusp catastrophe are shown
in figure~\ref{CuspFig} as a function of $a$. Again, they describe the
qualitative behavior of the bifurcating orbits close to the bifurcation.
It should be noted that the stationary values (\ref{CuspVal}) are negative
for all values of $a$, so that for a bifurcation described by (\ref{Cusp}),
the actions of the asymmetric orbits must be smaller than those of the
symmetric orbit. An alternative bifurcation scenario is described by the
dual cusp, viz.\ the
negative of (\ref{Cusp}). The dual
cusp is inequivalent to the regular cusp, but the scenario it describes
agrees with the above except that the stationary values and the second
derivatives change their signs, so that the actions of the asymmetric
orbits are now larger than that of the symmetric orbit.

\section{Closed orbits in the diamagnetic Kepler problem}
\label{sec:ClosedDKP}

As a basis for the description of closed orbits in the crossed-fields
hydrogen atom, we will choose the closed orbits in the diamagnetic Kepler
problem (DKP), i.e. in the hydrogen atom in a pure magnetic field. For
these orbits a complete classification is available
\cite{Main91,Main86,AlLaithy86,AlLaithy87,Main87,Mao92}. It will now be
recapitulated briefly.

For low scaled energies $\tilde E\to-\infty$, there are two fundamental
closed orbits: In one case, the electron leaves the nucleus parallel to the
magnetic field until the Coulomb attraction forces it back. This orbit is
purely Coulombic because the electron does not feel a Lorentz force when
moving parallel to the magnetic field. The second closed orbit lies in the
plane perpendicular to the magnetic field. Its shape is determined by the
combined influences of the Coulomb and magnetic fields.

Due to time-reversal invariance, both elementary orbits possess
arbitrary repetitions. As the scaled energy increases, each repetition of
an elementary orbit undergoes a sequence of bifurcations labelled by an
integer $\nu=1,2,3,\dots$ in order of increasing bifurcation energy. The
orbits born in these bifurcations can be characterized by the repetition
number $\mu$ of the bifurcating orbit and the bifurcation number
$\nu$. They are referred to \cite{Main91} as vibrators $V_\mu^\nu$ if they
bifurcate out of the orbit parallel to the magnetic field and as rotators
$R_\mu^\nu$ if they bifurcate out of the orbit perpendicular to $\vec B$.

\begin{figure}
  \includegraphics[width=\columnwidth]{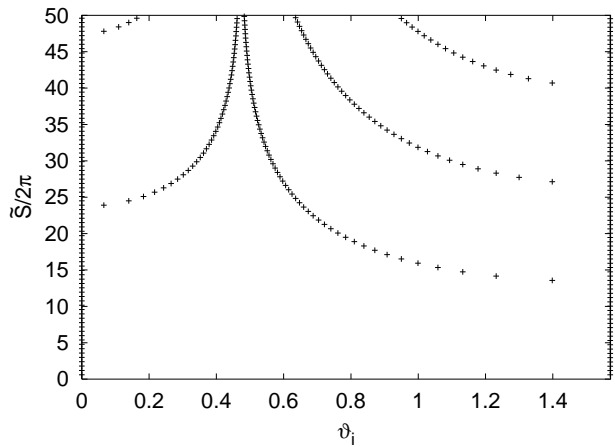}
  \caption{Scaled actions $\tilde S$ as functions of the  starting angles
  $\vartheta_i$ of closed orbits in the DKP for
  $\tilde E=-1.4$.}
  \label{DKPFig}
\end{figure}

Further bifurcations create additional orbits from the $V_\mu^\nu$ and
$R_\mu^\nu$ or ``exotic'' orbits not related to one of the two fundamental
orbits. These orbits are of importance at scaled energies
higher than those considered in this work, so that they will not be
discussed further. For the scaled energy $\tilde E=-1.4$, the scaled
actions and starting angles of the closed orbits are presented in
figure~\ref{DKPFig}. It can be seen that only orbits fitting into the
classification scheme described above are present. Furthermore, orbits
having a common bifurcation number $\nu$ lie on a smooth curve in the
plot. For this reason, we will refer to orbits characterized by a fixed
$\nu$ as a series of rotators or vibrators, respectively, and call $\nu$
the series number.

\section{Closed orbit bifurcation scenarios}
\label{sec:XCO}

In the presence of a pure magnetic field, the atomic system possesses a
rotational symmetry around the field axis. As a consequence, all closed
orbits except for the orbit parallel to the magnetic field occur in
continuous one-parameter families obtained by rotating a single orbit
around the symmetry axis. When a perpendicular electric field is
added, the rotational symmetry is broken. Out of each family, only two
orbits survive~\cite{Neumann97}, or, in other words, each family of orbits
splits into two independent orbits.

\subsection{Planar orbits}
\label{ssec:planar}

\begin{figure}
  \includegraphics[width=\columnwidth]{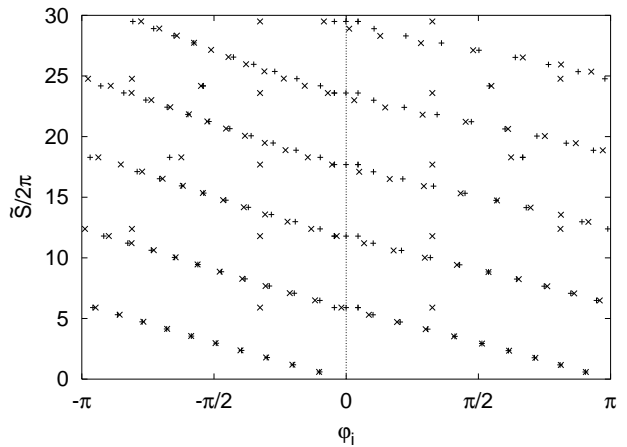}
  \caption{Scaled actions $\tilde S$ and azimuthal starting angles
  $\varphi_i$ for planar orbits at $\tilde E=-1.4$ and $\tilde F=0.03$ ($+$
  symbols) and $\tilde F=0.05$ ($\times$ symbols).}
  \label{Planar0Fig}
\end{figure}

The splitting of a family of orbits upon the introduction of an electric
field can most clearly be seen for planar orbits, i.e. for orbits lying in
the plane perpendicular to the magnetic field.  Due to the \textsf{Z}-symmetry,
this plane is invariant under the dynamics. Thus, the initial direction of
an orbit can be specified by means of the azimuthal angle $\varphi_i$ only.

Figure~\ref{Planar0Fig} shows the actions and initial directions of
the planar orbits for a scaled energy of $\tilde E=-1.4$ and scaled
electric field strengths $\tilde F=0.03$ and $\tilde F=0.05$. At $\tilde
F=0$, the orbits bifurcate off a certain repetition of the planar closed
orbit of the diamagnetic Kepler problem. For low $\tilde F$ they can 
therefore be assigned a repetition number. It can clearly be discerned in
figure~\ref{Planar0Fig} from the actions of the orbits.

\begin{figure}
  \includegraphics[width=.95\columnwidth]{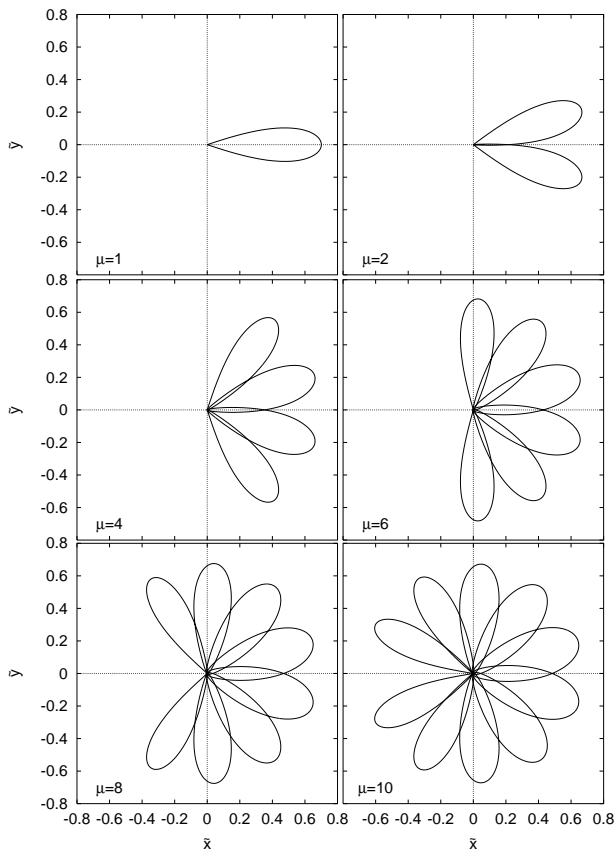}
  \caption{Elementary planar closed orbits at $\tilde E=-1.4$ and $\tilde
  F=0.05$. $\mu$ is the repetition number. The orbits are symmetric with
  respect to the $x$-axis, in particular, $\varphi_i=-\varphi_f$.}
  \label{PlanarTrajFig}
\end{figure}

As expected from the theory of the rotational symmetry breaking
\cite{Neumann97}, there are two orbits for each repetition number, and they
start in opposite directions from the nucleus. Moreover, the starting angle
varies linearly with the repetition number. These findings are illustrated
in figure~\ref{PlanarTrajFig}, where for a few low repetition numbers one
of the two orbits is shown. It can be seen that the orbits consist of more
and more ``loops'' and that the starting angle increases regularly. The
shapes are symmetric with respect to the $x$-axis because the orbits are
invariant under the \textsf{T}-transformation, i.e. these orbits are
singlets.

\begin{figure}
  \includegraphics[width=\columnwidth]{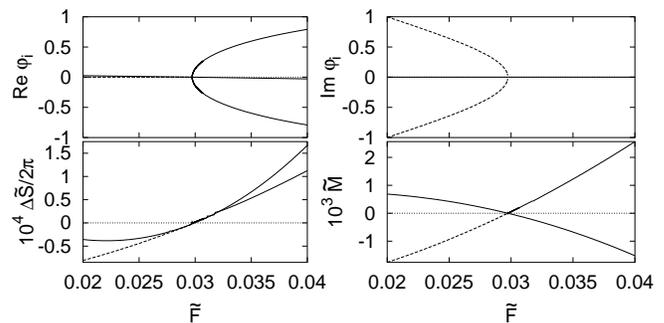}
  \caption{Orbital parameters 
  close to a pitchfork bifurcation creating a
  \textsf{Z}-doublet of closed orbits from a singlet at $\tilde E=-1.4$, repetition
  number $\mu=10$. 
  $\varphi_i$ is the azimuthal starting angle, $\tilde S$ the scaled action
  and $\tilde M$ the scaled stability determinant, $\Delta\tilde S=\tilde
  S-2\pi\times 5.898159$ was introduced for graphical purposes.
  Thick solid lines: singlet orbit, thin solid lines:
  doublet orbits, dashed line: ghost orbits symmetric with respect to
  complex conjugation. Dotted lines indicate coordinate axes.}
  \label{BifE10Fig}
\end{figure}

A few orbits in figure~\ref{Planar0Fig} do not fit into this simple
scheme. A closer inspection reveals that these orbits are not singlets, but
\textsf{Z}-doublets, and indeed they obviously occur in pairs. They are
generated by symmetry-breaking pitchfork bifurcations from singlet
orbits. Figure~\ref{BifE10Fig} presents the orbital parameters for closed
orbits involved in a bifurcation of this kind. The panels show the real and
imaginary parts of the starting angles $\varphi_i$ of the orbits, the
scaled actions $\tilde S$ and the stability determinant $M=\det (\partial
q_f/\partial p_i)$ (see section~\ref{sec:BifGen}), whose zeros indicate the
occurrence of a bifurcation and that also plays an important role in
closed-orbit theory \cite{Bartsch03b}.  These plots should be compared to
figure~\ref{CuspFig}, which displays the scenario described by the
symmetric cusp catastrophe. The qualitative agreement between the
catastrophe theory predictions and the numerical findings is evident.

A closer look at the asymmetric orbits reveals that they have equal initial
and final azimuthal angles $\varphi_i=\varphi_f$, i.e. they are not only
closed, but also periodic orbits. 
The initial and final angles of these orbits satisfy
$\varphi_i^{(1)}=-\varphi_f^{(2)}$ because they are symmetry partners and
$\varphi_i^{(2)}=\varphi_f^{(2)}$ because they are periodic. Thus, they
must fulfil $\varphi_i^{(1)}=-\varphi_i^{(2)}$. At the bifurcation, the
initial angles of the two orbits must coincide, so that a bifurcation can
only take place when $\varphi_i=0$ or $\varphi_i=\pi$, and it actually does
take place every time one of these conditions is fulfilled. This process
can be seen in figure~\ref{Planar0Fig}, e.g., at $S/2\pi\approx 25$: At
$\tilde F=0.03$, the symmetric orbit has not yet crossed the line
$\varphi_i=\pi$, so that no bifurcation has occurred. At $\tilde F=0.05$,
this line has been crossed and two asymmetric orbits have been created.

\begin{figure}
  \includegraphics[width=.95\columnwidth]{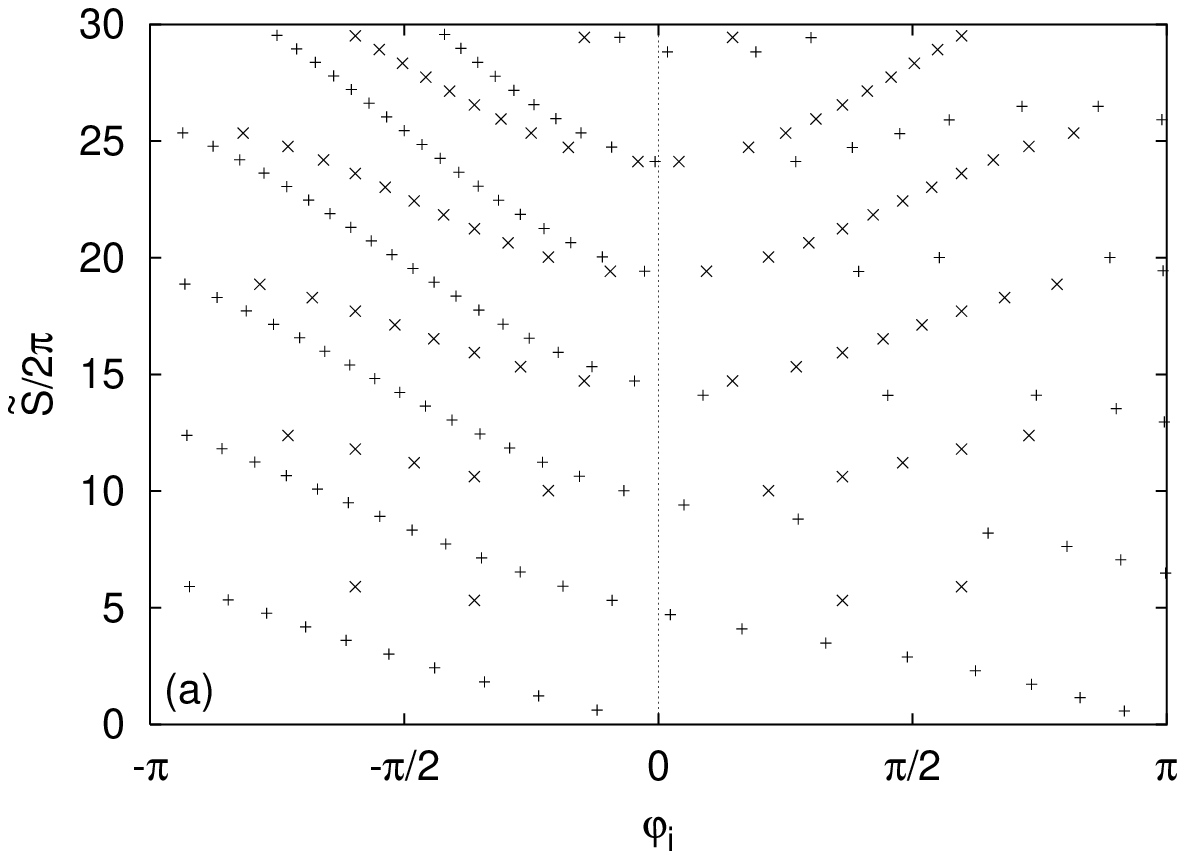}
  \includegraphics[width=.95\columnwidth]{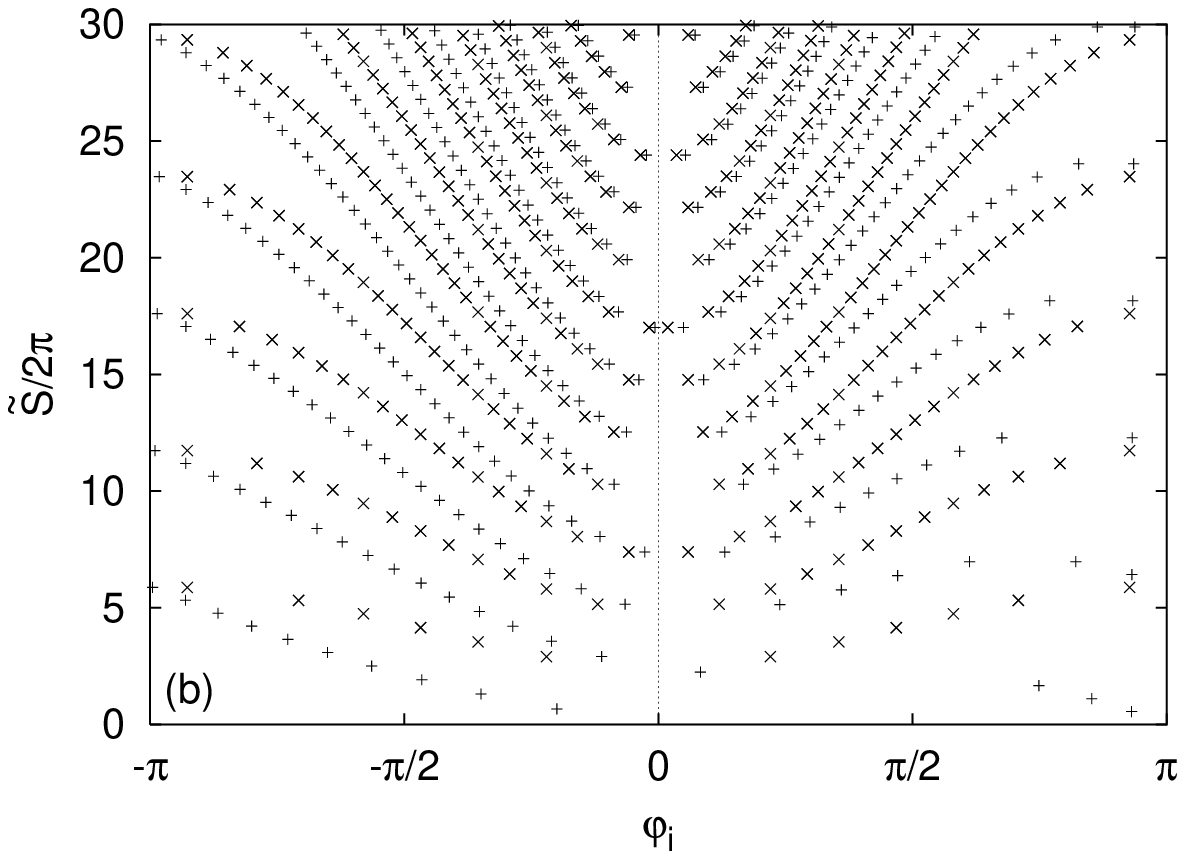}
  \caption{Scaled actions and azimuthal starting angles for planar orbits at
  $\tilde E=-1.4$ and (a) $\tilde F=0.2$, (b) $\tilde F=0.5$.
   Singlets are indicated by $+$ symbols,
   \textsf{Z}-doublets by $\times$ symbols.}
  \label{PlanarFig}
\end{figure}

\begin{figure}
  \includegraphics[width=\columnwidth]{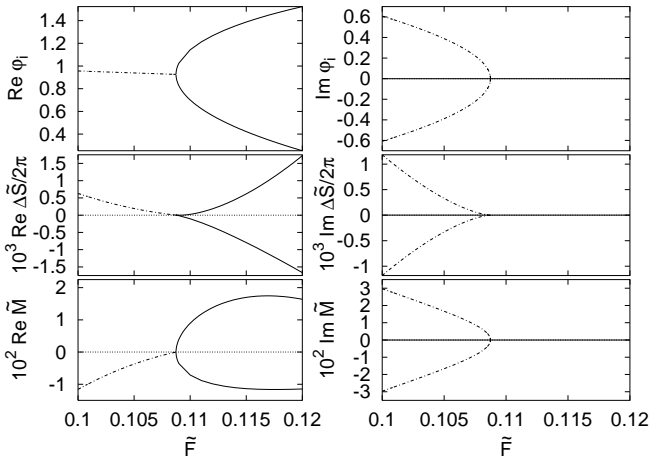}
  \caption{Orbital parameters close to a tangent bifurcation of planar
  orbits at a $\tilde E=-1.4$ and a winding number of
  $\mu=45$. ($\Delta\tilde S=\tilde S-2\pi\times26.512735$.) Solid lines:
  real orbits, dashed-dotted lines: ghost orbits.}
  \label{BifE45Fig}
\end{figure}

As the electric field strength is increased, the dependence of the starting
angle on the repetition number ceases to be linear. Instead, the curves
interpolating the functions $\tilde S(\varphi_i)$ start to develop humps,
so that at certain values of $\tilde S$, i.e. at certain repetition
numbers, more than two possible values of $\varphi_i$ exist. This
development is illustrated in figure~\ref{PlanarFig}. The humps indicate
the occurrence of tangent bifurcations generating additional pairs of
singlet orbits. This is the type of bifurcation described by the fold
catastrophe (\ref{Fold}). Orbital parameters for orbits involved in a
bifurcation of this kind are shown in figure~\ref{BifE45Fig}. As for the
pitchfork bifurcation, a comparison of that figure to the catastrophe
theory predictions in figure~\ref{FoldFig} reveals that the bifurcation is
well described qualitatively by the fold catastrophe.

\begin{figure}
  \includegraphics[width=\columnwidth]{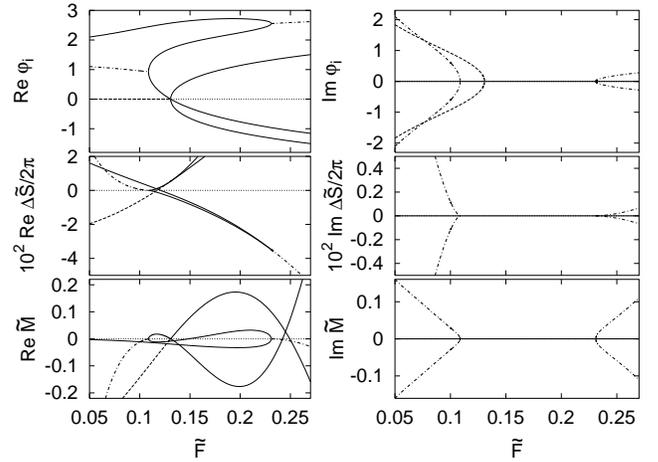}
  \caption{The bifurcation scenario taking place in the neighborhood of
  the tangent bifurcation shown in figure~\ref{BifE45Fig}. ($\Delta\tilde S=
  \tilde S-2\pi\times 26.2735$). Thick solid
  lines: real singlet orbits, thin solid lines: real \textsf{Z}-doublet orbits,
  dashed lines: symmetric ghost orbits,
  dashed-dotted lines: asymmetric ghost orbits. Dotted lines are coordinate
  axes.}
  \label{BifE45aFig}
\end{figure}

Once additional singlet orbits have been generated in a tangent
bifurcation, doublet orbits can be generated by pitchfork bifurcations in
the same way as from the original singlet orbits, i.e. a bifurcation will
occur whenever a singlet orbit crosses one of the lines $\varphi_i=0$ or
$\varphi_i=\pi$. This is illustrated in figure~\ref{BifE45aFig}, which
presents the tangent bifurcation already shown in figure~\ref{BifE45Fig} at
$\tilde F\approx0.11$. At $\tilde F=0.135$, one of the orbits thus
generated crosses the line $\varphi_i=0$, and two doublet orbits are
created from it. Together, the two bifurcations form what Wang and Delos
\cite{Wang01} call the ``normal sequence'' of bifurcations, whereas a
pitchfork bifurcation of a singlet orbit generated at $\tilde F=0$, which
is not preceded by a tangent bifurcation, is called a ``truncated series''.
These authors introduce an integrable model Hamiltonian to explain why this
kind of sequences can often be observed for planar orbits. The bifurcation
theory of sections~\ref{sec:BifGen} and~\ref{sec:Cod1Bif} sheds new light
on this question, suggesting that normal sequences can actually be expected
to occur even more generally than surmised by Wang and Delos. In
particular, although the crossed-fields system is close to integrable at
the field strengths considered here, integrability is not needed to make
pitchfork bifurcations a generic phenomenon. Instead, the presence of a
reflection symmetry suffices to reduce its codimension from two to one. The
sequence of a tangent and a pitchfork bifurcation, represented as a
sequence of a fold and a symmetric cusp catastrophe, can be regarded as an
unfolding of the symmetrized version of the butterfly catastrophe
\cite{Saunders80,Main97a}
\begin{equation}
  \label{SymmButt}
  \Phi_{a,b}(t) = \frac 16 t^6 - \frac 14 a t^4 - \frac 12 b t^2 \;,
\end{equation}
which is of codimension two, so that its unfolding can be expected to occur
frequently in codimension one.

A third bifurcation can be discerned in figure~\ref{BifE45aFig}: At $\tilde
F\approx 0.225$, a singlet orbit generated at $\tilde F=0$ and a singlet
orbit generated in the tangent bifurcation discussed above collide and are
destroyed. This is an instant of an inverse tangent bifurcation, which can
be described by the fold catastrophe in the same way as the ``regular''
tangent bifurcation. It forms the third building block for the bifurcation
scenario changing the pattern of planar orbits as the electric field
strength is increased.

Besides the three bifurcations described above, in figure~\ref{BifE45aFig}
three further zeros of the stability determinant $\tilde M$ occur for
certain real orbits, indicating the presence of even more
bifurcations. These bifurcations involve non-planar orbits, i.e. they are
pitchfork bifurcations breaking the \textsf{Z}-symmetry all planar orbits
possess. They will be discussed further in subsequent sections.  At the
moment it suffices to note that in this scenario six individual
bifurcations take place in a comparatively small interval of the electric
field strength. This is the first example of a phenomenon to be encountered
repeatedly: In the crossed-fields hydrogen atom bifurcations of closed
orbits abound, exacerbating both the classical and the semiclassical
treatment of the dynamics.

\subsection{Non-planar orbits}
\label{ssec:non-planar}

\begin{figure}
  \includegraphics[width=\columnwidth]{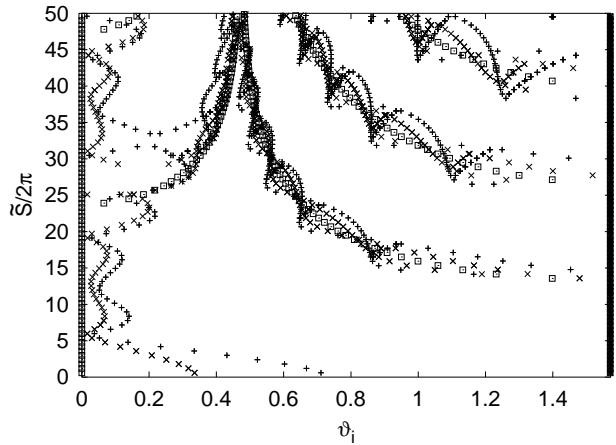}
  \caption{Scaled actions and polar starting angles of closed orbits at
  $\tilde E=-1.4$ and $\tilde F=0$ (pure magnetic field, $+$ symbols),
  $\tilde F=0.05$ ($\times$ symbols) and 
  $\tilde F=0.1$ ($\boxdot$ symbols). Due to the \textsf{Z}-symmetry, the
  figure should be extended to be symmetric with respect to the line
  $\vartheta_i=\pi/2$.}
  \label{RotSymmFig}
\end{figure}

The transition from the rotationally-symmetric case of a pure magnetic
field to crossed fields occurs for non-planar orbits in much the same way
as for planar orbits. As soon as a small perpendicular electric field is
present, a one-parameter family of DKP orbits is destroyed and splits into
two isolated closed orbits. These orbits start in opposite directions with
respect to the electric field, so that their azimuthal starting angles
$\varphi_i$ differ by $\pi$, in complete analogy with what was shown in
figure~\ref{Planar0Fig}. An additional complication arises because the
polar starting angle $\vartheta_i$ is no longer bound to the fixed value
$\pi/2$, so that the two orbits will in general have different
$\vartheta_i$. Figure~\ref{RotSymmFig} presents the polar starting angles
and the scaled actions of the closed orbits for the scaled energy $\tilde
E=-1.4$ in a pure magnetic field and for two different perpendicular
electric field strengths. Only angles $\vartheta_i\leq\pi/2$ need to be
shown because orbits with $\vartheta_i>\pi/2$ can be obtained by a
\textsf{Z}-reflection.  It is obvious from the figure how a family of
orbits splits in two isolated orbits and the two orbits move apart as the
electric field strength is increased. This process takes place in the same
way for both rotator and vibrator orbits.

An exceptional role is played by the DKP orbit parallel to the magnetic
field. This orbit is isolated even in a pure magnetic field. In the
presence of a perpendicular electric field it is distorted and torn away
from the magnetic field axis, but it remains isolated rather than splitting
into two orbits. This process is also apparent from
figure~\ref{RotSymmFig}. Notice again that closed orbits in crossed fields
do not possess repetitions. Any repetition of the parallel DKP orbit gives
rise to a closed orbit in crossed fields (for sufficiently small $\tilde
F$), but these orbits are not repetitions of each other. They have, in
particular, different starting angles.

The symmetries of the closed orbits are worth noting. All non-planar orbits
described so far are doublets. More precisely, the vibrator orbits are
\textsf{T}-doublets, i.e. they are invariant under the \textsf{T}\ operation. 
Their initial and final polar angles are small, as the orbits are mainly
directed along the magnetic field axis.

For the rotator orbits the situation is more complex. Their main component
is the motion in the plane perpendicular to the magnetic field, whereas the
$z$-component is comparatively small. They have, therefore, initial and
final polar angles close to $\pi/2$, so that it is conceivable that they
can start at an angle $\vartheta_i<\pi/2$ ``above'' the $x$-$y$-plane and
return at $\vartheta_f>\pi/2$ ``below'' that plane. This is in fact the
case for the rotators of the first series. They turn out to be \textsf{C}-doublets.

The second series of rotators contains orbits which, in the case of a pure
magnetic field, are composed of a first-series orbit and its \textsf{Z}-reflected
counterpart. The orbits of the second series therefore have
$\vartheta_i=\vartheta_f$ and are \textsf{T}-doublets. By the same token, orbits of
the third series return ``below'' the $x$-$y$-plane again and are
\textsf{C}-doublets, and higher series of rotators alternatingly contain
\textsf{T}-doublets and \textsf{C}-doublets.

\begin{figure}
  \includegraphics[width=.88\columnwidth]{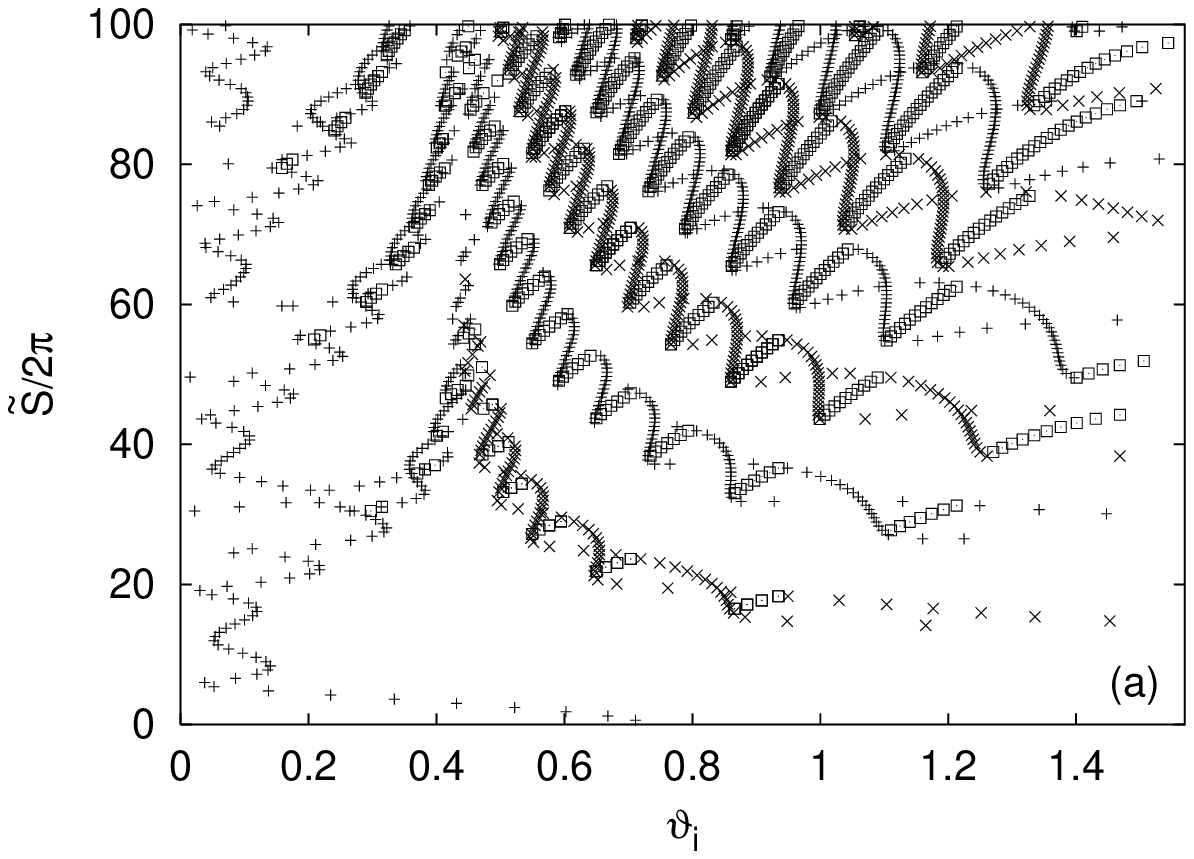}
  \includegraphics[width=.88\columnwidth]{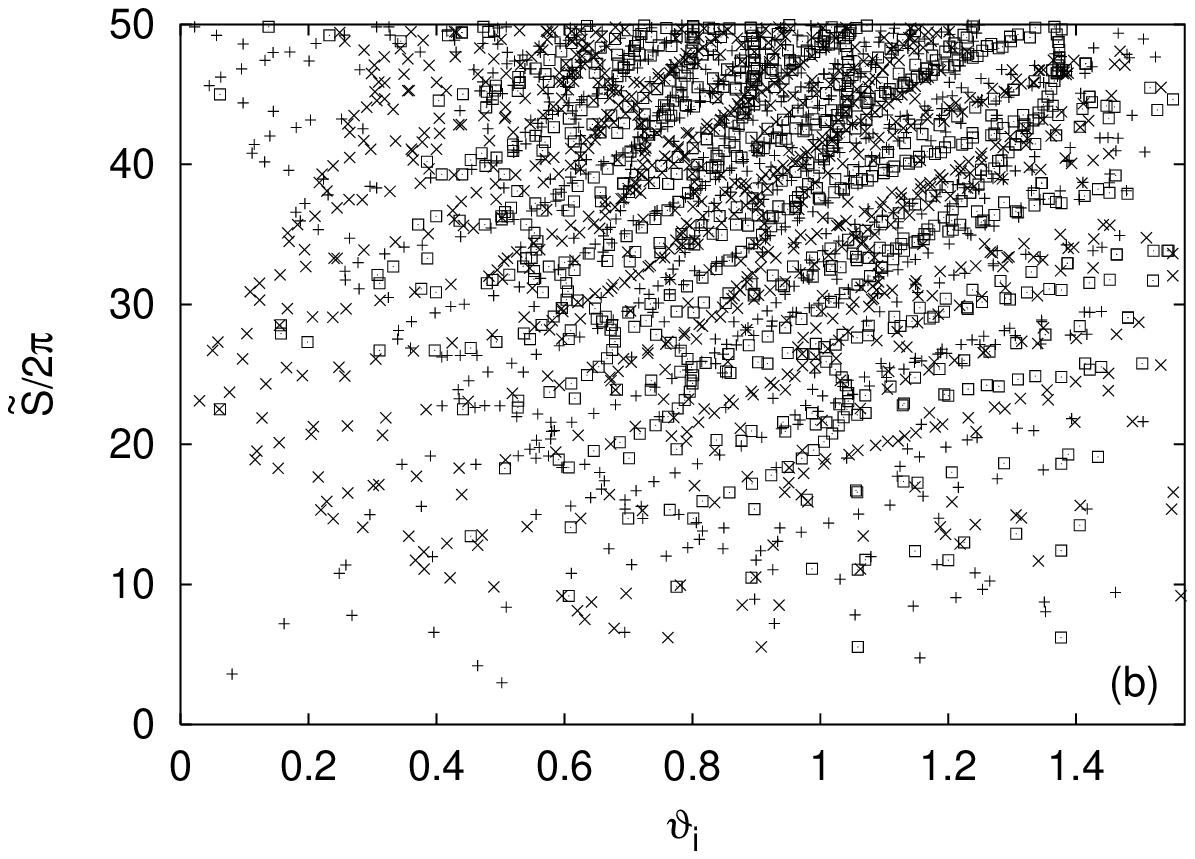}
  \caption{Scaled actions
  and polar starting angles of closed orbits at $\tilde E=-1.4$ and (a)
  $\tilde F=0.1$, (b) $\tilde F=0.6$.  Orbits are classified according to
  their symmetries: \textsf{T}-doublets are indicated by $+$~symbols,
  \textsf{C}-doublets
  by  $\times$~symbols, quartets by $\boxdot$~symbols.
  Planar orbits (\textsf{Z}-doublets and
  singlets) are omitted.  Notice that the range of actions shown is smaller
  in (b).}
  \label{DiscSymmFig}
\end{figure}

The distribution of symmetries is illustrated in
figure~\ref{DiscSymmFig}(a).  It extends the data given in
figure~\ref{RotSymmFig} to longer orbits and classifies the orbits
according to their symmetries.

\begin{figure}
  \includegraphics[width=\columnwidth]{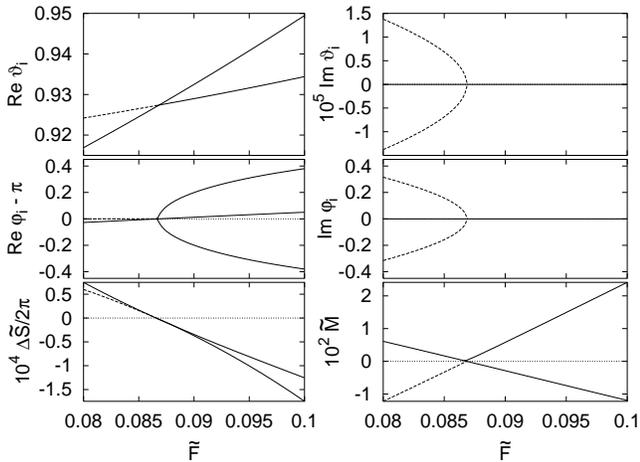}
  \caption{Orbital parameters close to a pitchfork bifurcation of a
  first-series 
  rotator and a repetition number of $\mu=38$. The
  bifurcation creates a quartet of orbits from a \textsf{C}-doublet.
  ($\Delta\tilde S=\tilde S-2\pi\times 18.297822$.)}
  \label{BifR38Fig}
\end{figure}

So far, only orbits present at arbitrarily low electric field strengths
have been described. As the electric field strength increases, further
bifurcations occur. Their general pattern can be identified in
figures~\ref{RotSymmFig} and~\ref{DiscSymmFig}(a). The most obvious
consequence of the bifurcations is the appearance of quartet orbits in each
series of both rotator and vibrator orbits. They are generated by pitchfork
bifurcations from the adjacent doublet orbits. As figure~\ref{BifR38Fig}
reveals if it is compared to figure~\ref{BifE10Fig}, this bifurcation is
very similar to a pitchfork bifurcation of planar orbits. A difference
arises because, due to the absence of \textsf{Z}-symmetry, the angle $\vartheta_i$
is not restricted to a fixed value.  As the \textsf{C}-symmetry to be broken
concerns the azimuth angles, it still is predominantly the angle
$\varphi_i$ that shows a square root behavior at the bifurcation and
obtains an imaginary part when ghost orbits exist. Nevertheless the polar
angle $\vartheta_i$ also acquires a small imaginary part. The real part of
$\vartheta_i$ apparently behaves linear close to the bifurcation, although
for electric field strengths above the critical value a square root
behavior must be present. It is too small to be seen in the figure. Even
though in quantitative terms the $\vartheta$-direction is only marginally
involved in the bifurcation, its presence has the important consequence
that the quartet orbits are no longer constrained to be periodic. As the
distance from the bifurcation is increased, the periodicity condition
$\varphi_i=\varphi_f$ is increasingly, albeit slowly, violated. The same
features can be found for the bifurcations introducing the quartet orbits
into the vibrator series.

\begin{figure}
  \includegraphics[width=\columnwidth]{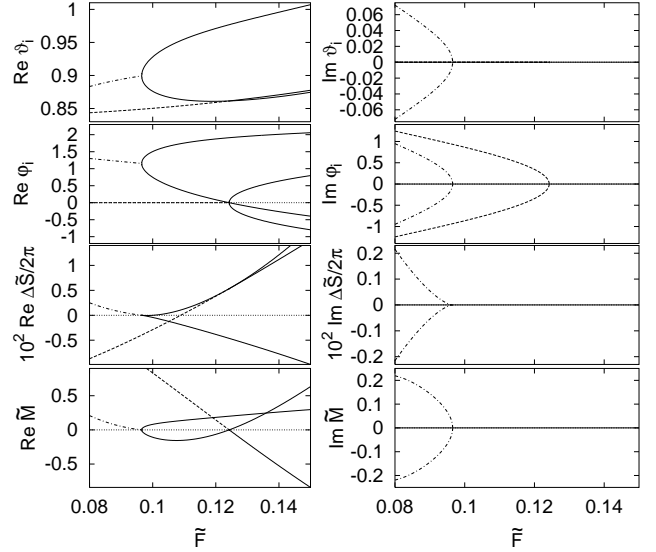}
  \caption{``Normal sequence'' of bifurcations for non-planar rotator
  orbits of the second series and a repetition number of $\mu=54$.
  ($\Delta\tilde S=\tilde S-2\pi\times 31.84035$.)}
  \label{BifR54Fig}
\end{figure}

The second important type of bifurcations is a tangent bifurcation
introducing new doublet orbits into the series. The occurrence of this
phenomenon can be noticed in figure~\ref{RotSymmFig}, if the numbers of
orbits of a given repetition number are compared for different electric
field strengths. An example of this bifurcation is given in
figure~\ref{BifR54Fig}. The tangent bifurcation involves both angles to
roughly equal extent. The two doublet orbits thus generated are implanted
into the regular pattern of their series, so that one of them subsequently
undergoes a pitchfork bifurcation which creates a quartet. This phenomenon
is entirely analogous to the ``normal sequence'' of bifurcations that was
found for planar orbits, except that the quartet orbits thus generated are
not periodic.

\begin{figure}
  \includegraphics[width=\columnwidth]{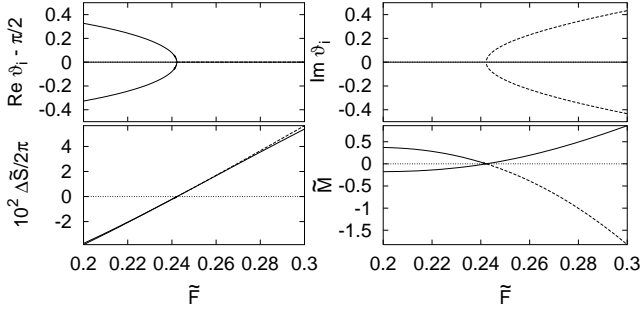}
  \caption{Destruction of \textsf{T}-doublet orbits in a collision with a singlet
  orbit. ($\Delta\tilde S=\tilde S-2\pi\times 27.60324$.)}
  \label{BifTEFig}
\end{figure}

\begin{figure}
  \includegraphics[width=\columnwidth]{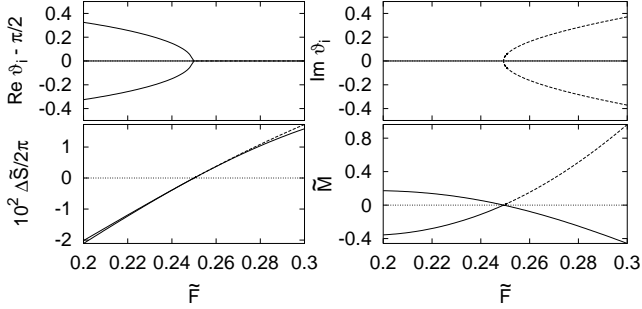}
  \caption{Destruction of quartet orbits in a collision with  \textsf{Z}-doublet
  orbits. ($\Delta\tilde S=\tilde S-2\pi\times 26.569655$.)}
  \label{BifQEFig}
\end{figure}

As the electric field strength increases, the rotator orbits of a given
series are torn apart and span an ever wider interval of
$\vartheta_i$. Those orbits moving towards higher values of $\vartheta_i$
eventually hit the plane $\vartheta_i=\pi/2$, where they collide with their
\textsf{Z}-reflected partners and are destroyed. One might suspect the
destruction of the two orbits to occur in a tangent bifurcation, but from
the discussion of section~\ref{sec:Cod1Bif} it is clear that a tangent
bifurcation can only create or destroy orbits having different actions, so
that it can never involve two orbits related by a symmetry
transformation. Thus, the bifurcation must be of pitchfork type, and it
must involve a \textsf{Z}-symmetric planar orbit.
Depending on whether the non-planar orbits colliding with the plane
are doublets or quartets, the planar orbit must be a singlet or a
\textsf{Z}-doublet, respectively. If the destruction scenario is regarded
in the direction of decreasing field strengths, it appears as the creation
of orbits with broken \textsf{Z}-symmetry from an orbit possessing this
symmetry. It is therefore the \textsf{Z}-breaking analogue of the
\textsf{T}- and \textsf{C}-symmetry breaking bifurcations described above.
As this type of bifurcation involves a planar orbit, it must give rise to a
zero in the stability determinant $\tilde M$ of the planar orbit. In fact,
the examples given in figures~\ref{BifTEFig} and~\ref{BifQEFig} for both
the destruction of a doublet and a quartet are two of the three
bifurcations whose presence was inferred from figure~\ref{BifE45aFig} the
discussion of the planar orbits.

The scenario just described is not restricted to rotator orbits. As can be
seen in figure~\ref{RotSymmFig}, the short vibrator orbits can, even at low
electric field strengths, reach rather high values of $\vartheta_i$. At
$\tilde F=0.15550$ the first of them collides, at $\vartheta_i=\pi/2$,
with its \textsf{Z}-reflected counterpart and is annihilated. This is a
pitchfork bifurcation in which one of the planar orbits with repetition
number $\mu=1$ takes part. Similarly, longer vibrators are destroyed in
collisions with planar rotators of the appropriate repetition numbers. This
example demonstrates that the distinction between vibrators and rotators,
which was borrowed from the case of vanishing electric field, does not
apply, strictly speaking, if an electric field is present. Although it is
generally useful for rather high electric field strengths, it can fail in
some instances. This is clearly the case when a bifurcation involves both
vibrator and rotator orbits.

\begin{figure}
  \includegraphics[width=\columnwidth]{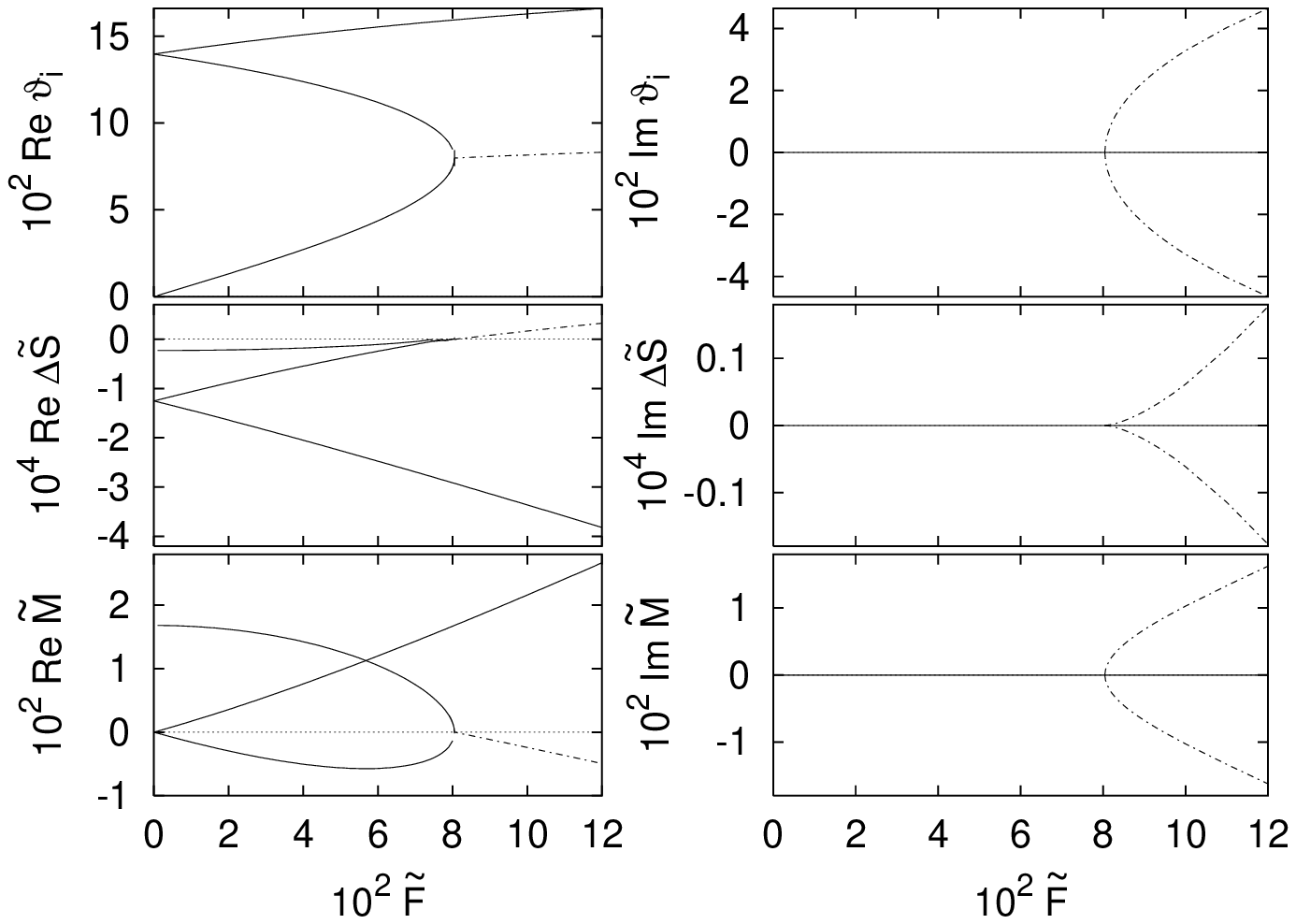}
  \caption{Simple bifurcation scenario for vibrator orbits of repetition
  number $\mu=41$. ($\Delta\tilde
  S=\tilde S-2\pi\times24.50221$.)}
  \label{BifV41Fig}
\end{figure}

A collision with the plane perpendicular to the magnetic field occurs only
for vibrators of low repetition numbers, and only for vibrators that
descend from the orbit parallel to the magnetic field. For longer orbits,
the usual scenario is different. At low electric field strength there is,
for sufficiently high repetition numbers, one orbit stemming from the orbit
parallel to the magnetic field and one or several pairs of orbits created
from non-parallel vibrators. It can be seen in figure~\ref{RotSymmFig},
however, that for certain repetition numbers two of these orbits can be
missing. This happens when the descendant of the parallel orbit and one of
the other vibrators annihilate in a tangent bifurcation. A simple example
of how this can come about is provided by the orbits with the repetition
number $\mu=41$. Their bifurcations are illustrated in
figure~\ref{BifV41Fig}. Two of the orbits obviously bifurcate from a common
family at $\tilde F=0$, whereas the orbit proceeding from the parallel
orbit is isolated there and starts at $\vartheta_i=0$. It then merges with
one of the other orbits in a tangent bifurcation to form a pair of ghost
orbits.

\begin{figure}
  \includegraphics[width=\columnwidth]{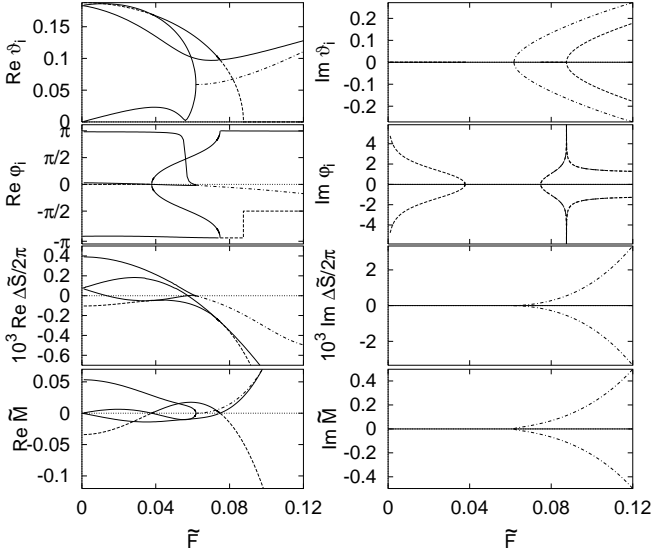}
  \caption{Complicated bifurcation scenario for vibrator orbits of repetition
  number $\mu=42$. ($\Delta\tilde
  S=\tilde S-2\pi\times 25.09941$.)}
  \label{BifV42Fig}
\end{figure}

This bifurcation is as simple as one could expect. For the neighboring
vibration number $\mu=42$ the scenario is more complicated. It is
illustrated in figure~\ref{BifV42Fig}. In this case, one of the orbits
generated in the rotational symmetry breaking at $\tilde F=0$, which is a
\textsf{T}-doublet, undergoes a pitchfork bifurcation and gives birth to a quartet
of orbits before it annihilates with the descendant of the parallel
orbit. The quartet then collides with the third, leftover
\textsf{T}-doublet and is destroyed in a second pitchfork
bifurcation.

Corresponding to the three elementary bifurcations, there are three ghost
orbits involved in the scenario. For one of them, the starting angles
$\vartheta_i$ and $\varphi_i$ show a peculiar behavior at the electric
field strength $\tilde F_0=0.08750$: Whereas $\vartheta_i$ exhibits a
square root behavior, changing from nearly real to nearly imaginary
values, the real part of $\varphi_i$ changes discontinuously by $\pi/2$,
and the imaginary part of $\varphi_i$ seems to diverge. Neither the action
nor the stability determinant of the orbit, on the contrary, show any kind
of special behavior. In particular, $\tilde M$ is non-zero, so that there
cannot be a bifurcation of the ghost orbit.

\begin{figure}
  \includegraphics[width=\columnwidth]{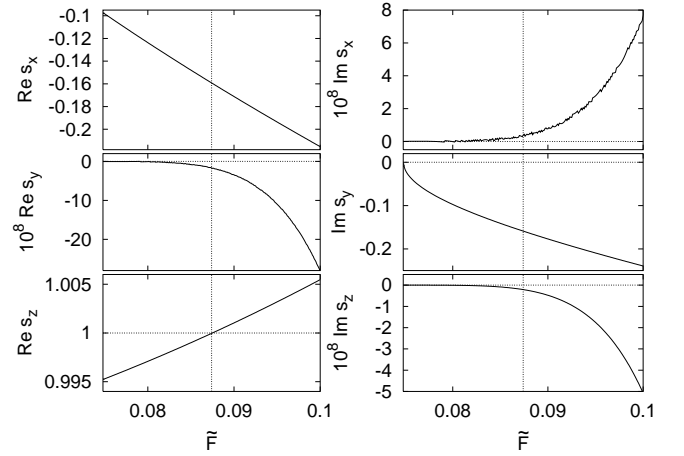}
  \caption{Cartesian components of the unit vector $\vec s$ specifying the
  starting direction of the ghost
  orbit ($s_x=\sin\vartheta_i\,\cos\varphi_i$,
  $s_y=\sin\vartheta_i\,\sin\varphi_i$, $s_z=\cos\vartheta_i$). Vertical
  dotted lines mark the field strength $\tilde F_0=0.08750$ where the
  singularity of $\Im\varphi_i$ is encountered.}
  \label{CartesianFig}
\end{figure}

The Cartesian components of the unit vector $\vec s$ in the starting
direction are given in figure~\ref{CartesianFig}. For all of them either
the real or the imaginary parts are small, so that their numerical
calculation is hard. Nevertheless, to within the numerical accuracy all
components are smooth at $\tilde F_0$, although the angles $\vartheta_i$
and $\varphi_i$ used to calculate them are not. Thus, the singularity must
be due to the transformation from Cartesian to angular coordinates. In the
real case it is obvious that the $(\vartheta,\varphi)$ coordinate chart is
singular at $\vartheta=0$.  To elucidate the details in the case of ghost
orbits, assume a model situation where $s_z=\cos\vartheta_i$ is exactly
real and $s_z=1$ at $\tilde F=\tilde F_0$. For ghost orbits, $s_z$ is not
bound to be smaller than 1, so that generically, to first order in
$\varepsilon=\tilde F-\tilde F_0$, $\cos\vartheta_i-1 \propto
\varepsilon$. Therefore, $\vartheta_i\propto\sqrt{\varepsilon}$ shows a
square root behavior and changes from purely real to purely imaginary
values. At the same time, $\sin\vartheta_i\propto\sqrt{\varepsilon}$ has a
zero, so that for $s_x=\sin\vartheta_i\,\cos\varphi_i$ and
$s_y=\sin\vartheta_i\,\sin\varphi_i$ to assume finite values,
$\sin\varphi_i$ and $\cos\varphi_i$ must diverge as
$\varepsilon^{-1/2}$. This is only possible if the imaginary part of
$\varphi_i$ is large. More precisely, if $\Im\varphi_i>0$ is assumed to be
large, $\sin\varphi_i$ and $\cos\varphi_i$ are proportional to
$e^{-i\varphi_i}$, whence
\begin{equation}
  \varphi_i=\frac{1}{2i}\ln\varepsilon 
              + {\cal O}\left(\varepsilon^0\right)
\end{equation}
achieves the desired divergence of $\sin\varphi_i$ and $\cos\varphi_i$.
Now $\Re\ln\varepsilon = \ln|\varepsilon|$ diverges at $\varepsilon=0$,
whereas $\Im\ln\varepsilon$ changes discontinuously from 0 to $\pm\pi$,
depending on what branch of the logarithm is chosen. This behavior results
in the observed divergence of $\Im\varphi_i$ and a discontinuous jump in
$\Re\varphi_i$ of size $\pi/2$.

\begin{figure}
  \includegraphics[width=\columnwidth]{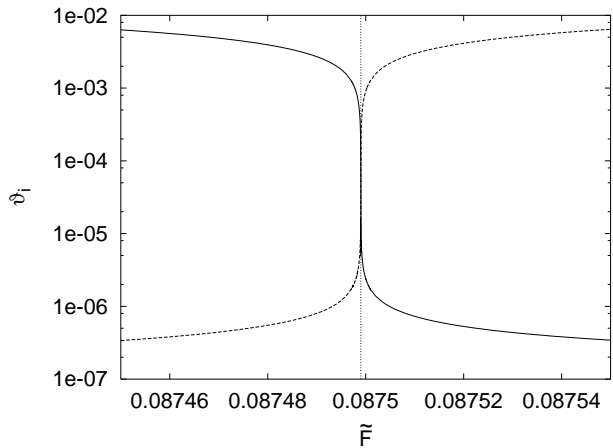}
  \caption{The starting angle $\vartheta_i$ of the ghost orbit close to the
  coordinate singularity. Solid line: real part, dashed line: imaginary
  part. Note the logarithmic scale.}
  \label{ThetaiFig}
\end{figure}

In the actual scenario $s_z$ will not be exactly equal to~1 at $\tilde F_0$
because this is a situation of real codimension two. However, if $\Im s_z$
is small, the singular behavior described above will be closely
approximated. Indeed, a closer look at $\vartheta_i$ (see
figure~\ref{ThetaiFig}) reveals that it is actually smooth, but close to
$\tilde F=\tilde F_0$ it changes extremely rapidly from almost real to
predominantly imaginary values. Similarly, the real part of $\varphi_i$ is
smooth, although it changes over an even smaller range of $\tilde F$. From
the numerical data it cannot be determined if $\Im\varphi_i$ is also smooth
or actually diverges. From the above discussion it is clear that it must be
smooth, because the coordinate singularity at $\vartheta_i=0$ is not
actually encountered.

It should be noted that the singularity described here can occur for ghost
orbits only. In the real case, as the pole $\vartheta_i=0$ on the real unit
sphere (which still has codimension two) is approached, both $s_x$ and
$s_y$ must vanish instead of assuming finite values, so that no divergences
of any kind are required.

\section{The classification of closed orbits}
\label{sec:class}

The fundamental classification scheme used in the above description of
closed orbit bifurcations is the distinction between rotators and
vibrators. This distinction was adopted from the case of vanishing electric
field strength, so it can be expected to be applicable if the electric
field is not too strong. In fact, from figure~\ref{DiscSymmFig}(a) it is
obvious that for $\tilde E=-1.4$ and $\tilde F=0.1$ orbits can uniquely be
classified as rotators or vibrators and can be assigned both a series
number and a repetition number. However, for long orbits neighboring
series of rotators already start to overlap in the plot, and if the
electric field strength is increased to $\tilde F=0.6$, all orbits get
completely mixed up, producing the somewhat messy picture shown in
figure~\ref{DiscSymmFig}(b).
The relative strengths of the electric and magnetic fields can conveniently
be compared if, instead of scaling the magnetic field strength to $1$ as
described in section~\ref{sec:classHam}, the scaling properties of the
Hamiltonian are used to scale the energy to $E=-1$. For the present
parameter values,  the scaled field strengths are then
$\tilde B=0.60$ and $\tilde F=0.30$, so that the magnetic field is still
stronger than the electric field, although the latter is no longer
negligibly small.
Thus, the pattern of orbits should still be dominated by the magnetic
field, but
figure~\ref{DiscSymmFig}(b) suggests that there is no way to achieve a
classification. 

\begin{figure}
  \includegraphics[width=\columnwidth]{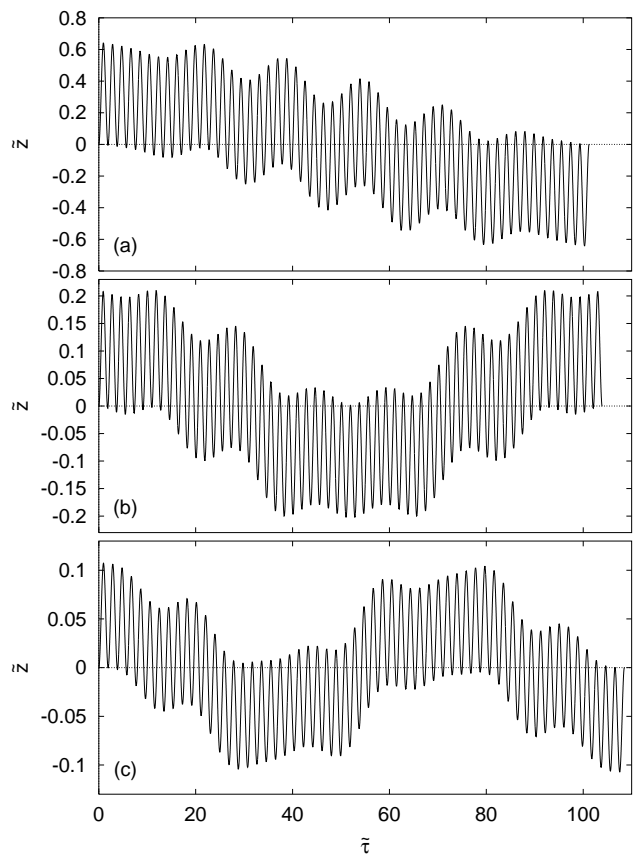}
  \caption{Rotor orbits of the (a) first, (b) second and (c) third series:
  scaled coordinate $\tilde z$ as a function of the scaled pseudotime
  $\tilde\tau$ for $\tilde E=-1.4$
  and $\tilde F=0.2$.}
  \label{RotFig}
\end{figure}

\begin{figure}
  \includegraphics[width=\columnwidth]{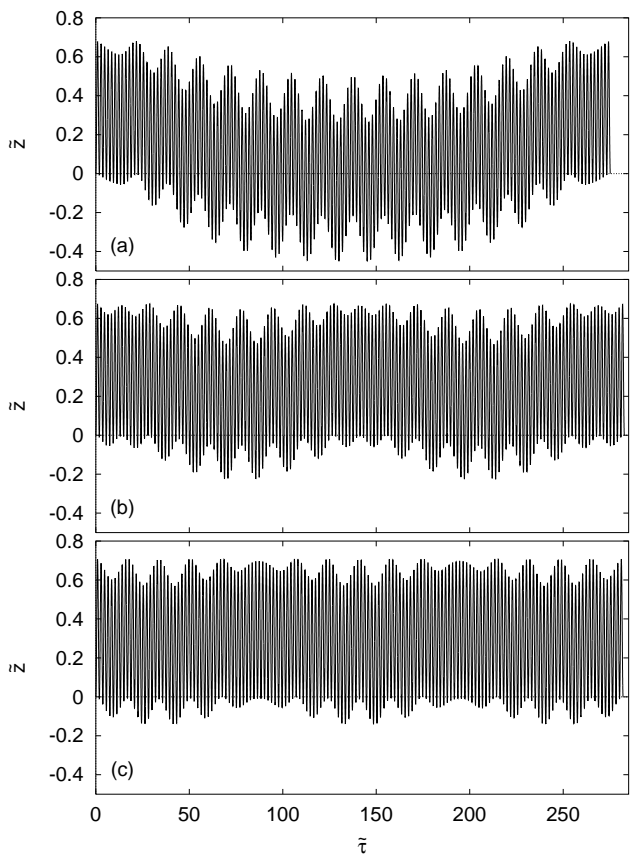}
  \caption{Vibrator orbits of the (a) first, (b) second and (c) third series:
  scaled coordinate $\tilde z$ as a function of the scaled pseudotime
  $\tilde \tau$ for $\tilde E=-1.4$
  and $\tilde F=0.2$.}
  \label{VibFig}
\end{figure}

A more suitable starting point for a classification is provided by the
complete trajectories. Of course, since the classification must gradually
break down for sufficiently strong electric fields, it can only be based on
heuristic criteria. The criteria we are going to propose are largely based on
the behavior of the $z$-coordinate of the motion as a function of time. To
illustrate them, this function is plotted for rotators of roughly equal
length from different series in figure~\ref{RotFig}. Figure~\ref{VibFig}
shows the analogous data for vibrators.

First of all, vibrators are connected to an orbit along the $z$-axis in the
pure magnetic field case. In this limit, the motion takes place either
``above'' the $x$-$y$-plane, i.e. in the half-space $z>0$, or ``below''
the plane. Rotators, on the contrary, stem from the elementary orbit in the
plane. Their motion takes place both above or below the plane. Rotators can
therefore be distinguished from vibrators if the maximum and minimum values
of the coordinate $z$ are compared: For a rotator, they must have roughly
equal absolute values, whereas for a vibrator ``above'' the plane, the
minimum value is much smaller in magnitude than the maximum value.

For the vibrators of a given length shown in figure~\ref{VibFig}, this
criterion gets better the higher the series of the vibrator is chosen. For
the vibrator of the first series, which is closest to the domain of
rotators, the excursion into the lower half space is of the same order of
magnitude as that into the upper half space. As the electric field
strength increases further, the vibrator orbit will eventually become
indistinguishable, by the present criteria, from a rotator of the second
series.

It has already been noted in the discussion of orbital symmetries that a
rotator of the first series that starts from the nucleus into the upper
half space returns to it from the lower half space, whereas a rotator of
the second series returns from the upper half space. This alternation
between motion in the upper and lower half spaces is obvious from
figure~\ref{RotFig}. It can be used to determine the series of a
rotator. If the value of $z(\tau)$ in a maximum is compared to its value in
the subsequent minimum, the transitions between motion in the upper and
lower half spaces can easily be monitored.

\begin{figure}
  \includegraphics[width=.8\columnwidth]{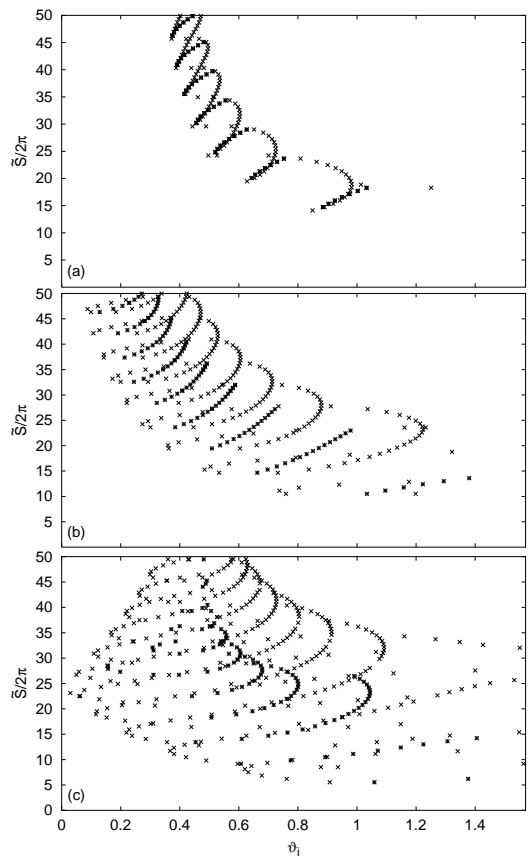}
  \caption{Rotators of the first series for $\tilde E=-1.4$ and (a) $\tilde
  F=0.2$, (b) $\tilde F=0.4$, and (c) $\tilde F=0.6$. \textsf{C}-doublets are
  indicated by $\times$~symbols, quartets by $\ast$~symbols.}
  \label{SeriesFig}
\end{figure}

Assigning a series number to a vibrator is considerably more difficult. It
relies on the beat structure present in $z(\tau)$ for a vibrator. Subsequent
maxima of this function have varying heights. We consider the maxima that
have minimal height compared to neighboring maxima, i.e. the minima of
the beats apparent in figure~\ref{VibFig}. The height of these beat minima
slowly oscillates, and the number of their oscillations turns out to give the
series number of the vibrator. It can be found by counting the number of
minima in the oscillation. Notice that the method uses extrema of the third
order: the minima in the minima in the height of maxima. It therefore
requires vibrators of sufficiently high repetition numbers, so that many
maxima of $z(\tau)$ exist. The method will fail when applied to vibrators
of very small repetition numbers. However, these vibrators exist for fairly
high scaled energies only, where even in the absence of an electric field
the dynamics is mixed or chaotic. In these regions, the classification
suggested here might not be meaningful at all.

Finally, orbits can be assigned a repetition number. For vibrators, this
can be done by simply counting the number of maxima in $z(\tau)$. For
rotators, maxima in $\rho(\tau)$ must be counted, where
$\rho=(x^2+y^2)^{1/2}$ is the distance from the magnetic field axis. In
this case, an additional difficulty arises because $\rho$ cannot assume
negative values, so that between two maxima corresponding to repetitions of
the elementary orbits, additional maxima arise from the ``swing-by''
around the nucleus. These maxima tend to be extremely small and
narrow close to the beginning and the end of an orbit, so that they are
hard to detect, but they can reach considerable heights in the middle of an
orbit. The present form of the classification algorithm, which is certainly
open to improvements, counts a maximum in $\rho(\tau)$ if its height
exceeds a certain fraction of the height of the previous maximum. In this
form, the assignment of rotator repetition numbers turns out to be the least
robust part of the classification procedure.

The criteria described above readily lend themselves to a numerical
implementation, so that the classification of orbits can be achieved
automatically. As an example, the rotators of the first series are shown in
figure~\ref{SeriesFig} for three different electric field
strengths. Although the neat ``wiggly-line'' structure characterizing the
series in figure~\ref{RotSymmFig} quickly breaks down for larger electric
field strengths, the distinction between different series
persists. Figure~\ref{SeriesFig}(c) should be compared to
figure~\ref{DiscSymmFig}(b). It might appear surprising that the
messy-looking set of orbits still permits a classification, but with the
help of the criteria just described an ordered pattern of closed orbits can
still be discerned. In this sense, the classification scheme derived from
the DKP turns out to be remarkably robust.

\section{Summary}

In this work, a systematic study of the closed classical orbits of the
hydrogen atom in crossed electric and magnetic fields has been carried
out. As an important step towards a complete understanding of the
complicated pattern of closed orbits, a bifurcation theory of closed orbits
was developed and the generic bifurcations of codimension one were
identified.

A variety of bifurcation scenarios observed in the crossed-fields system
was described. They demonstrate that, even though only two types of
elementary bifurcations exist, they combine into a wealth of complicated
bifurcation scenarios. The abundance of bifurcations exacerbates both a
complete classical description of the crossed-fields hydrogen atom and its
semiclassical treatment \cite{Bartsch03b}.

Based on the classification of closed orbits in the hydrogen atom in a
magnetic field, heuristic criteria have been proposed which allow a
systematization of closed orbits for moderately high electric field
strengths.  Although the present analysis cannot yet claim to have achieved
a complete classification of closed orbits in the crossed-fields hydrogen
atom, it does give a detailed impression of how orbits bifurcate as the
electric field strength increases. It thus introduces a high degree of
order into the complex set of closed orbits.


\begin{thebibliography}{31}
\expandafter\ifx\csname natexlab\endcsname\relax\def\natexlab#1{#1}\fi
\expandafter\ifx\csname bibnamefont\endcsname\relax
  \def\bibnamefont#1{#1}\fi
\expandafter\ifx\csname bibfnamefont\endcsname\relax
  \def\bibfnamefont#1{#1}\fi
\expandafter\ifx\csname citenamefont\endcsname\relax
  \def\citenamefont#1{#1}\fi
\expandafter\ifx\csname url\endcsname\relax
  \def\url#1{\texttt{#1}}\fi
\expandafter\ifx\csname urlprefix\endcsname\relax\def\urlprefix{URL }\fi
\providecommand{\bibinfo}[2]{#2}
\providecommand{\eprint}[2][]{\url{#2}}

\bibitem[{\citenamefont{Du and Delos}(1988)}]{Du88}
\bibinfo{author}{\bibfnamefont{M.~L.} \bibnamefont{Du}} \bibnamefont{and}
  \bibinfo{author}{\bibfnamefont{J.~B.} \bibnamefont{Delos}},
  \bibinfo{journal}{Phys.\ Rev.\ A} \textbf{\bibinfo{volume}{38}},
  \bibinfo{pages}{1896 and 1913} (\bibinfo{year}{1988}).

\bibitem[{\citenamefont{Bogomolny}(1989)}]{Bogomolny89}
\bibinfo{author}{\bibfnamefont{E.~B.} \bibnamefont{Bogomolny}},
  \bibinfo{journal}{Sov.\ Phys.\ JETP} \textbf{\bibinfo{volume}{69}},
  \bibinfo{pages}{275} (\bibinfo{year}{1989}).

\bibitem[{\citenamefont{Main}(1991)}]{Main91}
\bibinfo{author}{\bibfnamefont{J.}~\bibnamefont{Main}}, Ph.D. thesis,
  \bibinfo{school}{Universit{\"a}t Bielefeld, Germany} (\bibinfo{year}{1991}).

\bibitem[{\citenamefont{Main et~al.}(1986)\citenamefont{Main, Wiebusch, Holle,
  and Welge}}]{Main86}
\bibinfo{author}{\bibfnamefont{J.}~\bibnamefont{Main}},
  \bibinfo{author}{\bibfnamefont{G.}~\bibnamefont{Wiebusch}},
  \bibinfo{author}{\bibfnamefont{A.}~\bibnamefont{Holle}}, \bibnamefont{and}
  \bibinfo{author}{\bibfnamefont{K.~H.} \bibnamefont{Welge}},
  \bibinfo{journal}{Phys.\ Rev.\ Lett.} \textbf{\bibinfo{volume}{57}},
  \bibinfo{pages}{2789} (\bibinfo{year}{1986}).

\bibitem[{\citenamefont{Al-Laithy et~al.}(1986)\citenamefont{Al-Laithy,
  O'Mahony, and Taylor}}]{AlLaithy86}
\bibinfo{author}{\bibfnamefont{M.~A.} \bibnamefont{Al-Laithy}},
  \bibinfo{author}{\bibfnamefont{P.~F.} \bibnamefont{O'Mahony}},
  \bibnamefont{and} \bibinfo{author}{\bibfnamefont{K.~T.}
  \bibnamefont{Taylor}}, \bibinfo{journal}{J.\ Phys.\ B}
  \textbf{\bibinfo{volume}{19}}, \bibinfo{pages}{L773} (\bibinfo{year}{1986}).

\bibitem[{\citenamefont{Al-Laithy and Farmer}(1987)}]{AlLaithy87}
\bibinfo{author}{\bibfnamefont{M.~A.} \bibnamefont{Al-Laithy}}
  \bibnamefont{and} \bibinfo{author}{\bibfnamefont{C.~M.}
  \bibnamefont{Farmer}}, \bibinfo{journal}{J.\ Phys.\ B}
  \textbf{\bibinfo{volume}{20}}, \bibinfo{pages}{L747} (\bibinfo{year}{1987}).

\bibitem[{\citenamefont{Main et~al.}(1987)\citenamefont{Main, Holle, Wiebusch,
  and Welge}}]{Main87}
\bibinfo{author}{\bibfnamefont{J.}~\bibnamefont{Main}},
  \bibinfo{author}{\bibfnamefont{A.}~\bibnamefont{Holle}},
  \bibinfo{author}{\bibfnamefont{G.}~\bibnamefont{Wiebusch}}, \bibnamefont{and}
  \bibinfo{author}{\bibfnamefont{K.~H.} \bibnamefont{Welge}},
  \bibinfo{journal}{Z.\ Phys.\ D} \textbf{\bibinfo{volume}{6}},
  \bibinfo{pages}{295} (\bibinfo{year}{1987}).

\bibitem[{\citenamefont{Mao and Delos}(1992)}]{Mao92}
\bibinfo{author}{\bibfnamefont{J.-M.} \bibnamefont{Mao}} \bibnamefont{and}
  \bibinfo{author}{\bibfnamefont{J.~B.} \bibnamefont{Delos}},
  \bibinfo{journal}{Phys.\ Rev.\ A} \textbf{\bibinfo{volume}{45}},
  \bibinfo{pages}{1746} (\bibinfo{year}{1992}).

\bibitem[{\citenamefont{Raithel et~al.}(1991)\citenamefont{Raithel, Fauth, and
  Walther}}]{Raithel91}
\bibinfo{author}{\bibfnamefont{G.}~\bibnamefont{Raithel}},
  \bibinfo{author}{\bibfnamefont{M.}~\bibnamefont{Fauth}}, \bibnamefont{and}
  \bibinfo{author}{\bibfnamefont{H.}~\bibnamefont{Walther}},
  \bibinfo{journal}{Phys.\ Rev.\ A} \textbf{\bibinfo{volume}{44}},
  \bibinfo{pages}{1898} (\bibinfo{year}{1991}).

\bibitem[{\citenamefont{Raithel et~al.}(1994)\citenamefont{Raithel, Held,
  Marmet, and Walther}}]{Raithel94}
\bibinfo{author}{\bibfnamefont{G.}~\bibnamefont{Raithel}},
  \bibinfo{author}{\bibfnamefont{H.}~\bibnamefont{Held}},
  \bibinfo{author}{\bibfnamefont{L.}~\bibnamefont{Marmet}}, \bibnamefont{and}
  \bibinfo{author}{\bibfnamefont{H.}~\bibnamefont{Walther}},
  \bibinfo{journal}{J.\ Phys.\ B} \textbf{\bibinfo{volume}{27}},
  \bibinfo{pages}{2849} (\bibinfo{year}{1994}).

\bibitem[{\citenamefont{Rao et~al.}(2001)\citenamefont{Rao, Delande, and
  Taylor}}]{Rao01}
\bibinfo{author}{\bibfnamefont{J.}~\bibnamefont{Rao}},
  \bibinfo{author}{\bibfnamefont{D.}~\bibnamefont{Delande}}, \bibnamefont{and}
  \bibinfo{author}{\bibfnamefont{K.~T.} \bibnamefont{Taylor}},
  \bibinfo{journal}{J.\ Phys.\ B} \textbf{\bibinfo{volume}{34}},
  \bibinfo{pages}{L391} (\bibinfo{year}{2001}).

\bibitem[{\citenamefont{Freund et~al.}(2002)\citenamefont{Freund, Ubert,
  Fl{\"o}thmann, Welge, Wang, and Delos}}]{Freund02}
\bibinfo{author}{\bibfnamefont{S.}~\bibnamefont{Freund}},
  \bibinfo{author}{\bibfnamefont{R.}~\bibnamefont{Ubert}},
  \bibinfo{author}{\bibfnamefont{E.}~\bibnamefont{Fl{\"o}thmann}},
  \bibinfo{author}{\bibfnamefont{K.}~\bibnamefont{Welge}},
  \bibinfo{author}{\bibfnamefont{D.~M.} \bibnamefont{Wang}}, \bibnamefont{and}
  \bibinfo{author}{\bibfnamefont{J.~B.} \bibnamefont{Delos}},
  \bibinfo{journal}{Phys.\ Rev.\ A} \textbf{\bibinfo{volume}{65}},
  \bibinfo{pages}{053408} (\bibinfo{year}{2002}).

\bibitem[{\citenamefont{Gourlay et~al.}(1993)\citenamefont{Gourlay, Uzer, and
  Farrelly}}]{Gourlay93}
\bibinfo{author}{\bibfnamefont{M.~J.} \bibnamefont{Gourlay}},
  \bibinfo{author}{\bibfnamefont{T.}~\bibnamefont{Uzer}}, \bibnamefont{and}
  \bibinfo{author}{\bibfnamefont{D.}~\bibnamefont{Farrelly}},
  \bibinfo{journal}{Phys.\ Rev.\ A} \textbf{\bibinfo{volume}{47}},
  \bibinfo{pages}{3113} (\bibinfo{year}{1993}), \bibinfo{note}{erratum
  48,2508}.

\bibitem[{\citenamefont{Fl{\"o}thmann et~al.}(1994)\citenamefont{Fl{\"o}thmann,
  Main, and Welge}}]{Floethmann94}
\bibinfo{author}{\bibfnamefont{E.}~\bibnamefont{Fl{\"o}thmann}},
  \bibinfo{author}{\bibfnamefont{J.}~\bibnamefont{Main}}, \bibnamefont{and}
  \bibinfo{author}{\bibfnamefont{K.~H.} \bibnamefont{Welge}},
  \bibinfo{journal}{J.\ Phys.\ B} \textbf{\bibinfo{volume}{27}},
  \bibinfo{pages}{2821} (\bibinfo{year}{1994}).

\bibitem[{\citenamefont{von Milczewski and Uzer}(1997)}]{Milczewski97b}
\bibinfo{author}{\bibfnamefont{J.}~\bibnamefont{von Milczewski}}
  \bibnamefont{and} \bibinfo{author}{\bibfnamefont{T.}~\bibnamefont{Uzer}},
  \bibinfo{journal}{Phys.\ Rev.\ E} \textbf{\bibinfo{volume}{55}},
  \bibinfo{pages}{6540} (\bibinfo{year}{1997}).

\bibitem[{\citenamefont{Sadovski{\'\i} and
  Zhilinski{\'\i}}(1998)}]{Sadovskii98}
\bibinfo{author}{\bibfnamefont{D.~A.} \bibnamefont{Sadovski{\'\i}}}
  \bibnamefont{and} \bibinfo{author}{\bibfnamefont{B.~I.}
  \bibnamefont{Zhilinski{\'\i}}}, \bibinfo{journal}{Phys.\ Rev.\ A}
  \textbf{\bibinfo{volume}{57}}, \bibinfo{pages}{2867} (\bibinfo{year}{1998}).

\bibitem[{\citenamefont{Berglund and Uzer}(2001)}]{Berglund00}
\bibinfo{author}{\bibfnamefont{N.}~\bibnamefont{Berglund}} \bibnamefont{and}
  \bibinfo{author}{\bibfnamefont{T.}~\bibnamefont{Uzer}},
  \bibinfo{journal}{Found.\ Phys.} \textbf{\bibinfo{volume}{31}},
  \bibinfo{pages}{283} (\bibinfo{year}{2001}).

\bibitem[{\citenamefont{Wang and Delos}(2001)}]{Wang01}
\bibinfo{author}{\bibfnamefont{D.~M.} \bibnamefont{Wang}} \bibnamefont{and}
  \bibinfo{author}{\bibfnamefont{J.~B.} \bibnamefont{Delos}},
  \bibinfo{journal}{Phys.\ Rev.\ A} \textbf{\bibinfo{volume}{63}},
  \bibinfo{pages}{043409} (\bibinfo{year}{2001}).

\bibitem[{\citenamefont{Ku{\'s} et~al.}(1993)\citenamefont{Ku{\'s}, Haake, and
  Delande}}]{Kus93}
\bibinfo{author}{\bibfnamefont{M.}~\bibnamefont{Ku{\'s}}},
  \bibinfo{author}{\bibfnamefont{F.}~\bibnamefont{Haake}}, \bibnamefont{and}
  \bibinfo{author}{\bibfnamefont{D.}~\bibnamefont{Delande}},
  \bibinfo{journal}{Phys.\ Rev.\ Lett.} \textbf{\bibinfo{volume}{71}},
  \bibinfo{pages}{2167} (\bibinfo{year}{1993}).

\bibitem[{\citenamefont{Main and Wunner}(1997)}]{Main97a}
\bibinfo{author}{\bibfnamefont{J.}~\bibnamefont{Main}} \bibnamefont{and}
  \bibinfo{author}{\bibfnamefont{G.}~\bibnamefont{Wunner}},
  \bibinfo{journal}{Phys.\ Rev.\ A} \textbf{\bibinfo{volume}{55}},
  \bibinfo{pages}{1743} (\bibinfo{year}{1997}).

\bibitem[{\citenamefont{Bartsch et~al.}()\citenamefont{Bartsch, Main, and
  Wunner}}]{Bartsch03b}
\bibinfo{author}{\bibfnamefont{T.}~\bibnamefont{Bartsch}},
  \bibinfo{author}{\bibfnamefont{J.}~\bibnamefont{Main}}, \bibnamefont{and}
  \bibinfo{author}{\bibfnamefont{G.}~\bibnamefont{Wunner}},
  \bibinfo{note}{following paper}.

\bibitem[{\citenamefont{Mayer}(1970)}]{Mayer70}
\bibinfo{author}{\bibfnamefont{K.~R.} \bibnamefont{Mayer}},
  \bibinfo{journal}{Trans.\ AMS} \textbf{\bibinfo{volume}{149}},
  \bibinfo{pages}{95} (\bibinfo{year}{1970}).

\bibitem[{\citenamefont{Kustaanheimo and Stiefel}(1965)}]{KS65}
\bibinfo{author}{\bibfnamefont{P.}~\bibnamefont{Kustaanheimo}}
  \bibnamefont{and} \bibinfo{author}{\bibfnamefont{E.}~\bibnamefont{Stiefel}},
  \bibinfo{journal}{J.\ Reine Angew.\ Mathematik}
  \textbf{\bibinfo{volume}{218}}, \bibinfo{pages}{204} (\bibinfo{year}{1965}).

\bibitem[{\citenamefont{Bartsch}(2002)}]{Bartsch02}
\bibinfo{author}{\bibfnamefont{T.}~\bibnamefont{Bartsch}},
  \emph{\bibinfo{title}{The hydrogen atom in an electric field and in crossed
  electric and magnetic fields: Closed-orbit theory and semiclassical
  quantization.}} (\bibinfo{publisher}{Cuvillier},
  \bibinfo{address}{G\"ottingen, Germany}, \bibinfo{year}{2002}).

\bibitem[{\citenamefont{Bartsch}()}]{Bartsch03c}
\bibinfo{author}{\bibfnamefont{T.}~\bibnamefont{Bartsch}}, \bibinfo{note}{to be
  published}.

\bibitem[{\citenamefont{Goldstein}(1965)}]{Goldstein}
\bibinfo{author}{\bibfnamefont{H.}~\bibnamefont{Goldstein}},
  \emph{\bibinfo{title}{Classical Mechanics}}
  (\bibinfo{publisher}{Addison-Wesley Publishing Company},
  \bibinfo{address}{Reading,~MA}, \bibinfo{year}{1965}).

\bibitem[{\citenamefont{McDuff and Salamon}(1995)}]{McDuff95}
\bibinfo{author}{\bibfnamefont{D.}~\bibnamefont{McDuff}} \bibnamefont{and}
  \bibinfo{author}{\bibfnamefont{D.}~\bibnamefont{Salamon}},
  \emph{\bibinfo{title}{Introduction to symplectic topology}}
  (\bibinfo{publisher}{Clarendon Press}, \bibinfo{address}{Oxford},
  \bibinfo{year}{1995}).

\bibitem[{\citenamefont{Poston and Stewart}(1978)}]{Poston78}
\bibinfo{author}{\bibfnamefont{T.}~\bibnamefont{Poston}} \bibnamefont{and}
  \bibinfo{author}{\bibfnamefont{I.}~\bibnamefont{Stewart}},
  \emph{\bibinfo{title}{Catastrophe Theory and its Applications}}
  (\bibinfo{publisher}{Pitman}, \bibinfo{address}{Boston},
  \bibinfo{year}{1978}).

\bibitem[{\citenamefont{Saunders}(1980)}]{Saunders80}
\bibinfo{author}{\bibfnamefont{P.~T.} \bibnamefont{Saunders}},
  \emph{\bibinfo{title}{An introduction to catastrophe theory}}
  (\bibinfo{publisher}{Cambridge University Press},
  \bibinfo{address}{Cambridge, UK}, \bibinfo{year}{1980}).

\bibitem[{\citenamefont{Castrigiano and Hayes}(1993)}]{Castrigiano93}
\bibinfo{author}{\bibfnamefont{D.~P.~L.} \bibnamefont{Castrigiano}}
  \bibnamefont{and} \bibinfo{author}{\bibfnamefont{S.~A.} \bibnamefont{Hayes}},
  \emph{\bibinfo{title}{Catastrophe Theory}}
  (\bibinfo{publisher}{Addison-Wesley Publishing Company},
  \bibinfo{address}{Reading, MA}, \bibinfo{year}{1993}).

\bibitem[{\citenamefont{Neumann et~al.}(1997)\citenamefont{Neumann, Ubert,
  Freund, Fl{\"o}thmann, Sheehy, Welge, Haggerty, and Delos}}]{Neumann97}
\bibinfo{author}{\bibfnamefont{C.}~\bibnamefont{Neumann}},
  \bibinfo{author}{\bibfnamefont{R.}~\bibnamefont{Ubert}},
  \bibinfo{author}{\bibfnamefont{S.}~\bibnamefont{Freund}},
  \bibinfo{author}{\bibfnamefont{E.}~\bibnamefont{Fl{\"o}thmann}},
  \bibinfo{author}{\bibfnamefont{B.}~\bibnamefont{Sheehy}},
  \bibinfo{author}{\bibfnamefont{K.~H.} \bibnamefont{Welge}},
  \bibinfo{author}{\bibfnamefont{M.~R.} \bibnamefont{Haggerty}},
  \bibnamefont{and} \bibinfo{author}{\bibfnamefont{J.~B.} \bibnamefont{Delos}},
  \bibinfo{journal}{Phys.\ Rev.\ Lett.} \textbf{\bibinfo{volume}{78}},
  \bibinfo{pages}{4705} (\bibinfo{year}{1997}).

\end{thebibliography}

\end{document}